\documentclass[fleqn,usenatbib]{mnras}

\usepackage{newtxtext,newtxmath}

\usepackage[T1]{fontenc}

\DeclareRobustCommand{\VAN}[3]{#2}
\let\VANthebibliography\thebibliography
\def\thebibliography{\DeclareRobustCommand{\VAN}[3]{##3}\VANthebibliography}

\usepackage{graphicx}	
\usepackage{amsmath}	

\usepackage{amssymb}	

\title[] {Formation of metal-free binaries: Impact of H$_{2}$ line cooling and CIE cooling}

\author[R. Riaz,  D.R.G. Schleicher, S. Bovino, S. Vanaverbeke and Ralf S. Klessen]{R. Riaz$^{1,2}$\thanks{E-mail: rafeel.riaz@ubo.cl} 
	D.R.G. Schleicher$^{2}$\thanks{E-mail: dschleicher@astro-udec.cl}  
	S. Bovino$^{2}$\thanks{E-mail: stefanobovino@udec.cl}
	S. Vanaverbeke$^{3}$\thanks{E-mail: siegfriedvanaverbeke@gmail.com} Ralf S. Klessen$^{4,5}$\thanks{E-mail: klessen@uni-heidelberg.de} \\
	$^{1}$Centro de Investigación en Astronomía, Universidad Bernardo O'Higgins, Av. Viel 1497, Santiago, Región Metropolitana, Chile \\
	$^{2}$Departamento de Astronom\'ia, Facultad Ciencias F\'isicas y Matem\'aticas, Universidad de Concepci\'on, Av. Esteban Iturra s/n Barrio \\
	Universitario, Casilla $160$-C, Concepci\'on, Chile \\
	$^{3}$Centre for mathematical Plasma-Astrophysics, Department of Mathematics, KU Leuven, Celestijnenlaan 200B, 3001 Heverlee, Belgium \\ 
	$^{4}$Universit{\"a}t Heidelberg, Zentrum f{\"u}r Astronomie, Institut  f{\"u}r Theoretische Astrophysik, Albert-Ueberle-Str. 2,  69120 Heidelberg, Germany \\
	$^{5}$Universit{\"a}t Heidelberg, Interdisziplin{\"a}res Zentrum f{\"u}r Wissenschaftliches Rechnen, Im Neuenheimer Feld 205,  69120 Heidelberg, Germany
}

\date{Accepted XXX. Received YYY; in original form ZZZ}

\pubyear{2015}

\begin{document}
\label{firstpage}
\pagerange{\pageref{firstpage}--\pageref{lastpage}}
\maketitle

\begin{abstract}
During primordial star formation, the main cooling channel is provided by H$_{2}$ and super-molecules, such as H$_{2}$ or H$_{2}$, at sufficiently high densities. When the latter form at $n_{\rm H}$ $\geq$ $10^{14}$~cm$^{-3}$, collision-induced emission (CIE) provides efficient gas cooling. We investigate how CIE cooling affects the formation of metal-free binaries comparing simulations with and without this process. Irrespective of the cooling mechanism, we find a typical protostellar mass range between 0.01 to 100 M$_{\odot}$. However, models with only H$_{2}$ line cooling produce a greater number of low-mass protostars which exhibit stronger variations in their radial velocities than the high-mass protostars. Similarly, in models with both H$_{2}$ cooling and CIE cooling, significant variations in the radial velocities are found for protostars in the intermediate mass range. The initial number of fragments $N_{\rm max}$ decreases with increasing strength of turbulence. Cooling via super-molecules lets the most massive protobinaries (MMPBs) efficiently accrete mass. The maximum mass accretion rate $\dot M_{\rm max}$ for the MMPBs is more than an order of magnitude higher in the presence of CIE cooling than for pure H$_{2}$ line cooling. As a result, compact binaries with a semi-major axis as small as 3.57 au may form through the H$_{2}$ $-$ H$_{2}$ cooling channel. Our results indicate that in addition to the MMPBs most population III (Pop. III) binaries should be in eccentric i.e. non-circular orbits. This provides an important connection to the eccentric binaries reported in previous studies, which were found to exhibit rich temporal accretion signals during their evolution. 

\end{abstract}




\section{Introduction}
The formation of the first stars requires cooling in the primordial gas. According to the $\Lambda$CDM model, dark matter minihalos provide localized concentrations of otherwise diffuse primordial gas in the early universe \citep{bromm2013formation, glover2013first, ade2016planck}. However, for the gas to collapse to high densities, efficient cooling agents are required which play an important role in a metal-free environments. For the formation  of the first stars, it is now well established that molecular hydrogen (H$_{2}$) is one of the main cooling agents, which provides an efficient cooling mechanism for the primordial gas at $\lesssim$ 10$^{4}$ K \citep{saslaw1967molecular, matsuda1969cooling}. Studies related to the properties of H$_{2}$ have revealed that the temperatures inside the primordial gas in the presence of H$_{2}$ can drop down to $\simeq$ 200 K \citep{palla1983primordial, glover2008uncertainties,clark2011formation}. The temperature of the primordial gas is set primarily via H$_{2}$ cooling, allowing the gas to radiate away energy and the collapse to continue to form protostars. Within the gas cloud, individual fragments will become gravitationally unstable and start collapsing  if their mass is larger than the Jeans mass, and will subsequently form massive stars in the range of 10$-$1000 $M_\odot$ \citep{abel2002formation, bromm2004first,latif2015disc}. In addition, several high resolution numerical studies have indicated the formation of low-mass stars during the collapse of the primordial gas cloud, due to fragmentation occuring at very high densities \citep{clark2008first, greif2012formation, susa2013mass, stacy2014first,  prole2022fragmentation}.   

The gas density at which H$_{2}$ cooling operates via rovibrational line emission is $\sim 10^4$~cm$^{-3}$ \citep{abel1998first,abel2002formation, bromm2002formation}. As the density during the primordial gas collapse reaches $10^8$~cm$^{-3}$, three-body reactions come into play and give rise to a larger H$_{2}$ fraction \citep{palla1983primordial,bovino2014primordial}. The cooling in the primordial gas that is triggered by three-body reactions due to the release of the binding energy of the molecules is supported at densities between $10^{8}$~cm$^{-3}$ to $10^{10}$~cm$^{-3}$, and in fact, becomes the most efficient cooling mechanism from densities of $10^{10}$~cm$^{-3}$ \citep{yoshida2006formation}. However, as the gas collapses even further, the H$_{2}$ line cooling gets affected by the ever-increasing gas density which results in a higher opacity of the medium \citep{hirano2013radiative}. As a consequence, for densities 10$^{10}$ cm$^{-3}$ $\leq$~$n_{\rm H}$~$\leq$ $10^{14}$ cm$^{-3}$, the role of H$_{2}$ line cooling is significantly suppressed. The collision-induced emission (CIE) then takes over and starts to operate as another radiative coolant at densities $\geq$ $10^{14}$~cm$^{-3}$ \citep{omukai1998formation,ripamonti2004fragmentation, hirano2013radiative,glover2006cooling,clark2011gravitational}. 
As isolated H$_{2}$ molecules are symmetric, no dipole moment exists, thus the emission or absorption of radiation is only possible via quadruple transitions. However, at higher gas densities the chances of collisions among the H$_{2}$ molecules are much higher,  giving rise to the formation of a super-molecule \citep{ripamonti2004fragmentation, smith2007transition, clark2011gravitational, bromm2013formation, glover2013first, grassi2014krome, stacy2014mutual, van2014effects, sharda2019role}. This super-molecule forms a temporary dipole and makes the emission or absorption of radiation possible via dipole transitions  at gas densities of $\sim$ 10$^{14}$ cm$^{-3}$ where H$_{2}$ becomes less important as a cooling agent. In addition to the super-molecule H$_{2}$ $-$ H$_{2}$, H$_{2}$ can also form a supermolecule with H and He (i.e. H$_{2}$ $-$ He and H$_{2}$ $-$ H) \citep{ripamonti2004fragmentation}.  

Many of the numerical simulations performed to understand the fragmentation behaviour of  collapsing primordial gas and the resulting formation of  single, binary, and multiple stellar systems have introduced sink particles (protostars) at a density $n_{\rm H}$ $\leq$ $10^{13}$~cm$^{-3}$ \citep{bromm2004accretion, moeckel2010evolution, glover2010fragmentation, stacy2010first, hosokawa2011protostellar, stacy2011rotation, stacy2012first, susa2014mass, dutta2015role, dutta2016effects, riaz2018formation, sharda2019role, sugimura2020birth, latif2022birth}. This is prior to the density threshold $n_{\rm H}$ $\geq$ $10^{14}$~cm$^{-3}$ where CIE cooling comes into play. Hence, the formation and evolution of the Pop. III protostellar systems in these studies were subject to only the gas cooling provided by H$_{2}$ molecules. In this paper, our aim is to study the fragmentation behavior in the CIE cooling regime, including the impact on the formation of binaries, and we compare to simulations where the CIE cooling was not included.

In the first phase of the gas cooling i.e. during the H$_{2}$ line cooling, as explored by \citet{clark2011gravitational}, it is expected that turbulence at low Mach numbers allows fragmentation to occur and the forming clumps to survive. We, therefore, explore the primordial gas collapse at low Mach numbers to understand the process of fragmentation that may lead to the genesis of population III (Pop. III) binary systems \citep{turk2009formation}. Studies of zero-metallicity gas collapsing in isolation (i.e. without external radiation sources) have reported fragmentation to occur at densities $\sim$ 10$^{8}$ cm$^{-3}$ \citep{bromm1999forming}. Once the gas density reaches $\sim$ 10$^{14}$ cm$^{-3}$ or even higher, the gas enters the approximately adiabatic regime of collapse and the temperature of the gas becomes relatively higher as well \citep{loeb1994collapse,omukai2001primordial, omukai2005thermal, yoshida2006formation}. In general, studying both phases of primordial gas cooling (i.e. H$_{2}$ line cooling and the CIE cooling) and the respective subsequent nature of fragmentation is vital to comprehend the true origin and formation pathway(s) of the Pop. III stars, and binaries in particular. The mass distribution associated with each type of metal-free stellar configuration is important to understand the possible fate of the Pop. III stars, as quantified by \citet{heger2002nucleosynthetic}. The stellar fate may affect the primordial gas present in the surroundings and thus can leave imprints on the gas composition, which hosts the next phase of star-formation \citep{2003ApJ...591...38W,2008ApJ...674..644C,2013RvMP...85..809K}. Also, as discussed by \citet{mebane2020effects}, the indirect detection of Pop. III stars via their supernovae or their effect on the cosmological 21$-$cm background requires a better understanding of the binary configurations and the associated mass distributions along with other binary characteristics of the first stars. 

We present our simulation methodology in section 2. The initial conditions adopted for our simulations are described in section 3. Section 4 is reserved for the results and discussion which is followed by section 5 where we provide the conclusions and further outlook.

\section{Methods}\label{methods}
We employ the numerical code GRADSPH$-$KROME, which is a coupled version of the  
smoothed particle hydrodynamics (SPH) code GRADSPH\footnote{Webpage GRADSPH: http://www.swmath.org/software/1046} \citep{vanaverbeke2009gradsph} and the chemistry package KROME\footnote{Webpage KROME: http://www.kromepackage.org/} \citep{grassi2014krome}. This coupled numerical scheme allows us to simulate the hydrodynamics of the star-forming gas including chemistry and cooling as presented by \citet{riaz2018formation}, which we now refer to as RBVS hereafter. We use the  H$_{2}$ cooling function  provided  by  \citet{2008MNRAS.388.1627G} with the update from \citet{2015MNRAS.451.2082G}. We also include continuum and Compton cooling as described by \citet{2000ApJ...534..809O}, and consider the formation/destruction of H$_{2}$ and the associated energy sources/sinks that produce the heating/cooling of the gas. For the CIE cooling, we use the cooling function provided by \citet{ripamonti2004fragmentation}. We adopt the simple optical depth approximation employed by \citet{ripamonti2004fragmentation}, in spite of some limitations that were pointed out by \citet{hartwig2015new}. The simple optical depth approximation underestimates the photon escape probability as it neither take into account the density gradient nor the true shape of the cloud, implying that the real cooling could be somewhat more efficient than estimated here. 

In RBVS, a detailed discussion of the SPH scheme implemented in GRADSPH$-$KROME is provided including the sink particle formalism by \citet{hubber2013improved}. An additional feature of the most recent version of GRADSPH$-$KROME, which we use in this work, is the sink merger scheme by \citet{stacy2013constraining}. According to this scheme, a merger of two sinks is allowed when the following three criteria are satisfied:
	\begin{itemize}\setlength{\itemsep}{0.25cm}
		\item When their relative distance $d$ is smaller than the accretion radius $r_{\rm acc}$ so that $d < r_{\rm acc}$.
		\item When the total energy $E_{\rm tot}$ of the pair of sink particles is negative, so that the pair is gravitationally bound.
		\item When the least massive sink (secondary) of the pair has insufficient angular momentum to remain rotationally supported against infall onto the massive sink (primary) i. e.  $j_{\rm sec}$ $<$ $j_{\rm cent}$, where $j_{\rm cent}$=$\sqrt{G~M_{\rm primary}~d}$ and $M_{\rm primary}$ denotes the mass of the most massive sink of the pair. 
	\end{itemize} 
	
	For another more recent investigation that has used a similar merger scheme of sink particles, we refer to \citet{riaz2020fragmentation}. In order to determine pairs of sinks which form gravitationally bound binary systems, we define the total orbital energy per unit mass of a pair of sink particles as in \cite{stacy2013constraining}:
\begin{equation} \label{orbitalenergy}
\epsilon=\epsilon_{p}+\epsilon_{k},
\end{equation}
where $\epsilon_{p}$ and $\epsilon_{k}$ are the gravitational potential energy and kinetic energy per unit mass, respectively, and are defined as
\begin{equation} \label{potentialenergy}
\epsilon_{p}= -\frac{G\left(M_{1}+M_{2}\right)}{r},
\end{equation}
and
\begin{equation} \label{kineticenergy}
\epsilon_{k}=\frac{1}{2}v^{2},
\end{equation}

where $r$ is their mutual distance, $v$ is their relative velocity, and $M_{1}$ and $M_{2}$ are the two protostellar masses of the pair, respectively. A pair of sinks is considered a binary if $\epsilon < 0$. Sink particles which do not fulfil these criteria are treated as isolated protostars.

We terminate all of our simulations when the star formation efficiency (SFE) reaches 2 per cent. This stopping criterion is opted for due to the significant computational cost of each run. It further ensures that the simulations can be compared when they are in a similar stage, i.e. after the same fraction of mass has been converted into protostars, and thereby facilitates the comparison between the two sets of models, i.e. M1a$-$M5a and M1b$-$M5b.

\begin{table}
	\centering
	\caption{Summary of the initial physical parameters of the simulation models M1a$-$M5a and M1b$-$M5b. The table describes the turbulent Mach number ($\mathcal{M}$), the ratio of the thermal energy of the cloud to its gravitational potential energy ($\alpha_{\rm turb}$), the sink formation density threshold ($n_{\rm sink}$) and the accretion radius ($r_{\rm acc}$). For each model, the initial radius of the cloud, the total mass inside the cloud, the ratio of kinetic energy to the
gravitational potential energy of the gas cloud ($\alpha_{\rm th}$), the initial average number density of the cloud, and the initial temperature of the gas are given by the constant values $2.169$~pc, $1.3041 \times 10^{4}$~M$_{\odot}$, $0.196$, $8.650 \times 10^{4}$~cm$^{-3}$ and $300$~K, respectively.}
	\label{tab:Table1}
	\begin{tabular}{ccccc} 
		\hline
		\hline
		Model & $\mathcal{M}$ & $\alpha_{\rm turb}$ & $n_{\rm sink}$ (cm$^{-3}$) & $r_{\rm acc}$ (au)\\
		\hline
		M1a & 0.1 & 6.541$ \times 10^{-4}$ & 10$^{13}$ & 28\\
		M2a & 0.2 & 2.616$ \times 10^{-3}$ & 10$^{13}$ & 28\\
		M3a & 0.4 & 1.046$ \times 10^{-2}$ & 10$^{13}$ & 28\\
		M4a & 0.8 & 4.186$ \times 10^{-2}$ & 10$^{13}$ & 28\\
		M5a & 1.0 & 6.541$ \times 10^{-2}$ & 10$^{13}$ & 28\\
		\hline
		M1b & 0.1 & 6.541$ \times 10^{-4}$ & 10$^{15}$ & 5\\
		M2b & 0.2 & 2.616$ \times 10^{-3}$ & 10$^{15}$ & 5\\
		M3b & 0.4 & 1.046$ \times 10^{-2}$ & 10$^{15}$ & 5\\
		M4b & 0.8 & 4.186$ \times 10^{-2}$ & 10$^{15}$ & 5\\
		M5b & 1.0 & 6.541$ \times 10^{-2}$ & 10$^{13}$ & 5\\
		\hline
	\end{tabular}
\end{table}

\section{Initial conditions}
Our models consist of two sets M1a$-$M5a and M1b$-$M5b (see Table 1). The first set M1a$-$M5a is following the evolution of the gas in the H$_{2}$ line cooling regime and employs a density threshold for sink particle formation that becomes relevant before the densities of CIE cooling are being reached. In the second set, the primordial gas cloud is allowed to collapse to first pass through the H$_{2}$ line cooling phase and then continues to collapse until it enters the higher density regime ($\geq10^{14}$~cm$^{-3}$) where the CIE cooling mechanism is relevant. 

As initial condition, we assume a spherical primordial gas cloud consisting of 1150709 SPH particles, which remains identical to RBVS. The total mass inside the gas cloud is $M = 1.3041 \times 10^{4}$~M$_{\odot}$, the radius of the cloud is $R = 2.169$~pc, the initial gas density is $n_{\rm i} = 8.650 \times 10^{4}$~cm$^{-3}$, and the gas initially is at a temperature $T = 300$~K. The total mass inside the sphere is three times the Jeans mass of the system. The gas is under solid-body rotation. We use a rotational parameter $\beta{}$ = 5 per cent which is defined as the ratio of the rotational energy to the gravitational potential energy of the cloud. The gas is turbulent with the turbulent Mach number $\mathcal{M}$ set to $0.1$, $0.2$, $0.4$, $0.8$ and $1.0$ to explore the impact of different levels of turbulence. Due to the different values for the initial turbulent Mach number, we thus have five different models in each set denotes as M1a$-$M5a and M1b$-$M5b (see Table 1). The injection of a turbulent velocity spectrum in the initial conditions follows the same procedure previously adopted by RBVS. The spectral index $p$ of the initial turbulence in the present work is chosen to be 1.75, to mimic the gas that is compressible in nature. Our choice is inspired by the work of \citet{clark2011gravitational}, who adopted a turbulent velocity field that has a power spectrum slightly steeper than the standard description for incompressible flows \citep{kolmogorov1941equations} since the gas is compressible in nature. The ratio of the thermal energy ($U_{\rm th}$ = $\frac{1}{2}Mv^{2}$) to the gravitational potential energy ($\Omega$ = $-$ $\frac{3}{5} \frac{GM^{2}}{R}$) is described by the parameter $\alpha_{\rm th}$ defined as 
\begin{equation} \label{alpha_th}
	\alpha_{\rm th}=\frac{5 R k T}{2 G M \mu m_{\rm H}},
\end{equation}
where $R$, $k$ $T$, $G$, and $M$, $\mu$, $m_{\rm H}$ are the radius of the cloud, the Boltzmann constant, the initial gas temperature, the gravitational constant, the mass of the cloud, the mean molecular weight, and the mass of hydrogen atom, respectively. In our calculations, we set $\mu$ = 1.22 at the beginning. The parameter $\alpha_{\rm th}$ is set to 0.196 and remains identical in each of our models M1a$-$M5a and M1b$-$M5b. Similarly, the ratio of the turbulent energy ($U_{\rm turb}$ = $\frac{1}{2}M\mathcal{M}^{2}c^{2}$) to the gravitational potential energy $\Omega$ is described with a parameter $\alpha_{\rm turb}$ and defined as 
\begin{equation} \label{alpha_turb}
\alpha_{\rm turb}=\frac{U_{\rm turb}}{\mathbf{|\Omega|}}.
\end{equation}
The parameter $\alpha_{\rm turb}$ is model-dependent in each set (see Table 1). In our model simulations, the code uses $G=M=R=1$ as internal dimensionless units.

As we employ the same number of SPH particles as in RBVS and also take the number of neighbouring particles to be $N_{\rm opt}$ = 50 in each of our simulations, we have a mass resolution of $M_{\rm resolution}$ = 1.133 M$_{\odot}$ in the present work similar to RBVS, which is calculated as
	\begin{equation} \label{Mass Resolution}
	M_{\rm resolution}=2 N_{\rm opt} m_{\rm particle}.
	\end{equation}
We redefine the accretion radius $r_{\rm acc}$ of the sink particle in the present work, which takes a unique value in each set of our simulations. This is primarily due to the Jeans length that needs to be well-resolved in each simulation \citep{2010ApJ...717..121C,2011ApJ...731...62F}, which depends both on the gas density and the temperature during the collapse as \citep{tohline1980gravitational} 
\begin{equation} \label{Jeans Radius}
R_{\rm J}=0.76 \left(\frac{R_{\rm g}T}{G\rho}\right) ^{1/2},
\end{equation}
where $R_{\rm J}$, $R_{\rm g}$, $T$, $G$, and $\rho$, are the Jeans length, the gas constant, the temperature of the gas, the gravitational constant, and the mass density of gas, respectively. For a self-gravitating gas cloud, this relation describes the condition for an isolated lump of gas to collapse and form protostar(s) (see figure 3). If the lump has a characteristic size (radius) greater than the Jeans length $R_{\rm J}$ at a given density $\rho$  and temperature $T$, then the thermal pressure alone will not be able to prevent the gravitational collapse.

\begin{figure*}
	\includegraphics[angle=0,scale=0.425]{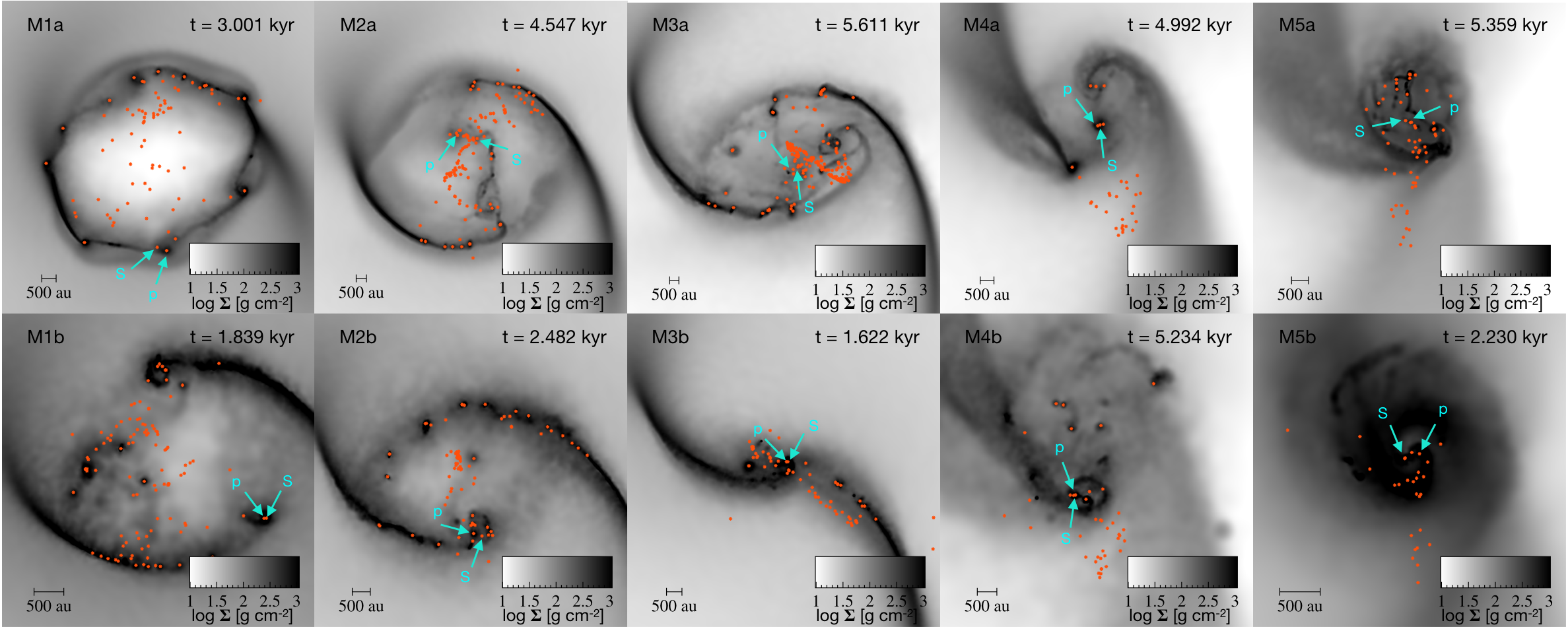}
	\caption{Simulation results for models M1a$-$M5a (top panels) and M1b$-$M5b (bottom panels) at the end of our computation when the SFE in each model reaches $\xi$ = 2 per cent. Each panel shows in logarithmic scale the projected column density ($ \Sigma$) in the xy-plane integrated along the z-axis in g cm$^{-2}$. Sink particles (protostars) are shown as orange dots. The arrows in each panel mark the primary (p) and secondary (s) components of the most massive protobinary (MMPB) located in the cluster. }
	\label{fig:figur6}
\end{figure*}

\begin{table}
	\centering
	\caption{Summary of the two sets of models M1a$-$M5a and M1b$-$M5b. All quantities are evaluated at the time when the SFE ($\xi$) reaches 2 per cent. The table describes the final time (since the formation of the first sink particle in the gas) of model termination ($t_{\rm final}$), the total number of protostars produced ($N_{\rm max}$), the final number of protostars after mergers ($N_{\rm proto}$), the binary fraction ($f_{\rm~binary}$) and the binary contribution towards the SFE ($\xi_{\rm binary}$).}
	\label{tab:Table2}
	\begin{tabular}{cccccc} 
		\hline
		\hline
		Model & $t_{\rm final}$ (kyr) & $N_{\rm max}$ & $N_{\rm proto}$  & $f_{\rm~binary}$ & $\xi_{\rm~binary}$\\
		\hline
		M1a & 3.001 & 81  & 17 & 0.500 & 0.017\\
		M2a & 4.547 & 128 & 30 &  0.366 & 0.042\\
		M3a & 5.611 & 191 & 34 &  0.699 & 0.041\\
		M4a & 4.992 & 27  & 6 &  0.625 & 0.041\\
		M5a & 5.359 & 58  & 25 &  0.166 & 0.012\\
		\hline
		M1b & 1.839 & 113 & 38 &  0.260 & 0.020\\
		M2b & 2.482 & 91  & 43 &  0.155 & 0.016\\
		M3b & 1.622 & 69  & 21 &  0.230 & 0.025\\
		M4b & 5.234 & 44  & 19 &  0.002 & 0.012\\
		M5b & 2.230 & 28  & 13 &  0.008 & 0.009\\
		\hline
	\end{tabular}
\end{table}

\begin{figure*}
	\includegraphics[angle=0,scale=0.425]{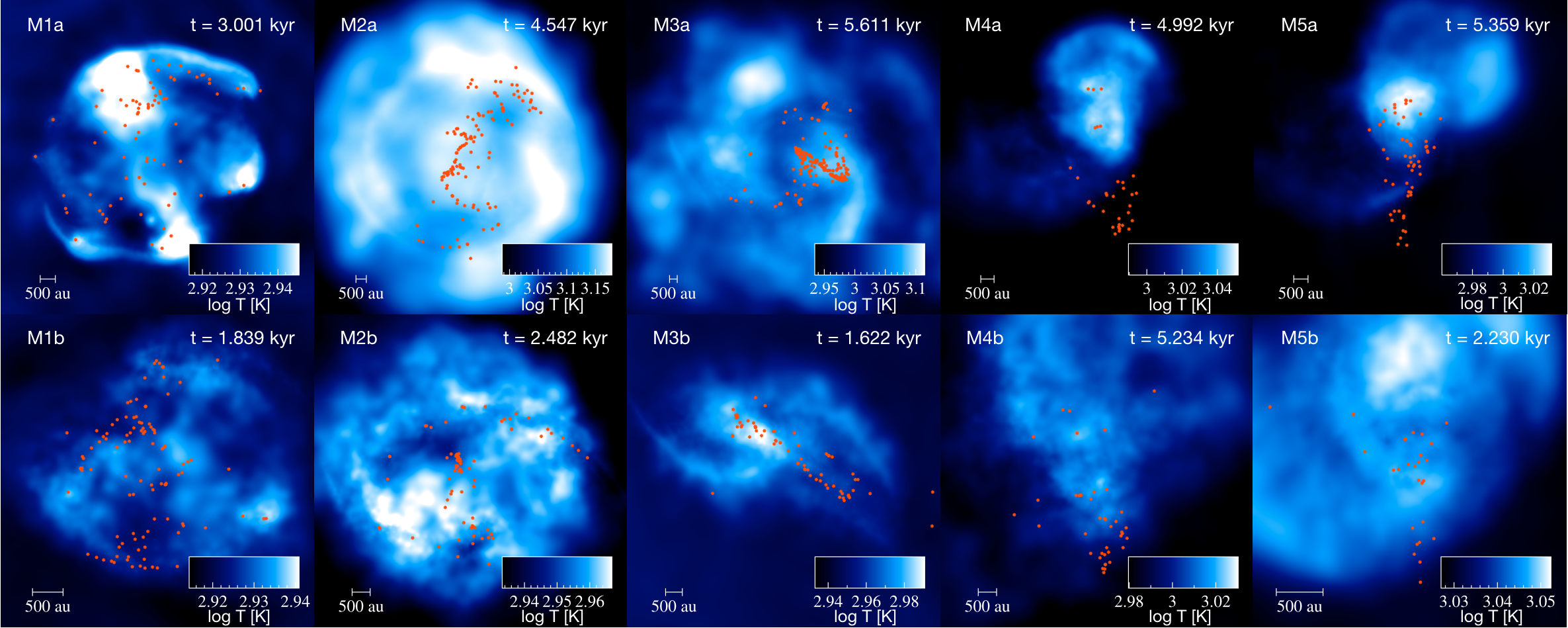}
	\caption{Simulation results for models M1a$-$M5a (top panels) and M1b$-$M5b (bottom panels) at the end of our simulations when the SFE in each model reaches $\xi$ = 2 per cent. Each panel (the xy-plane) shows in logarithmic scale the temperature ($T$) integrated along the z-axis in Kelvin. Sink particles (protostars) are shown as orange dots. }
	\label{fig:figur6}
\end{figure*}

\begin{figure*}
	\includegraphics[angle=0,scale=0.425]{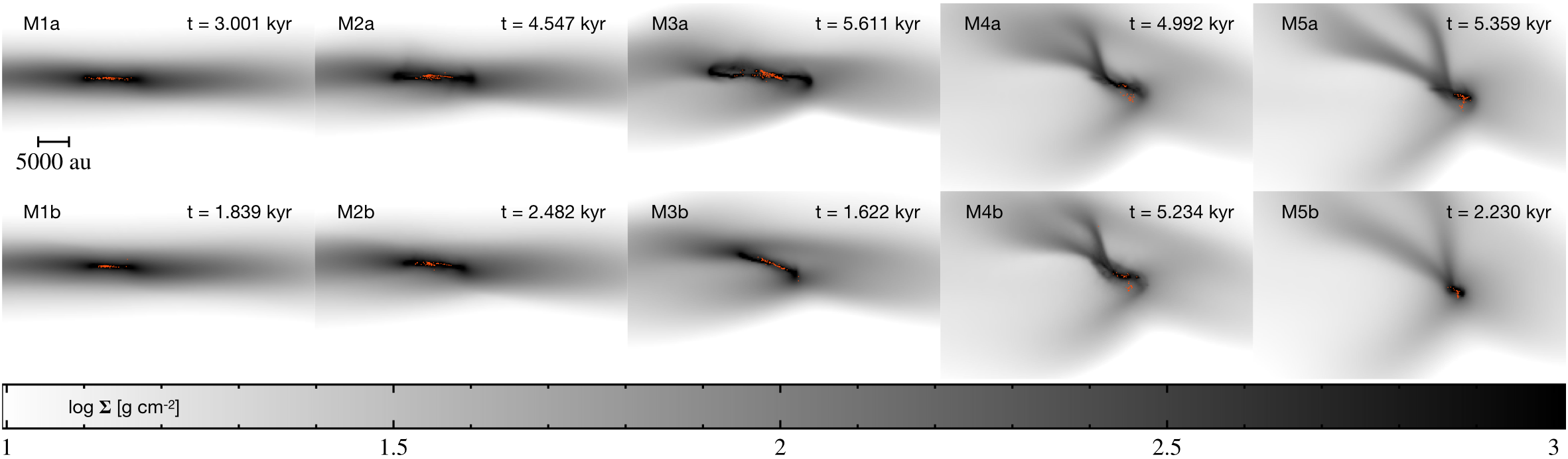}
	\caption{Global disc morphology for models M1a$-$M5a (top panels) and M1b$-$M5b (bottom panels) at the end of our simulations when the SFE in each model reaches $\xi$ = 2 per cent. In each row from left to right, an initial turbulent velocity field corresponding to $\mathcal{M}$ = 0.1 $-$ 1.0 is imposed onto solid-body rotation. Each panel (the xz-plane) shows in logarithmic scale the projected column density ($ \Sigma$) integrated along the y-axis in g cm$^{-2}$. Sink particles (protostars) are shown as orange dots. }
	\label{fig:figur6}
\end{figure*}

\begin{figure}
	\includegraphics[width=\columnwidth]{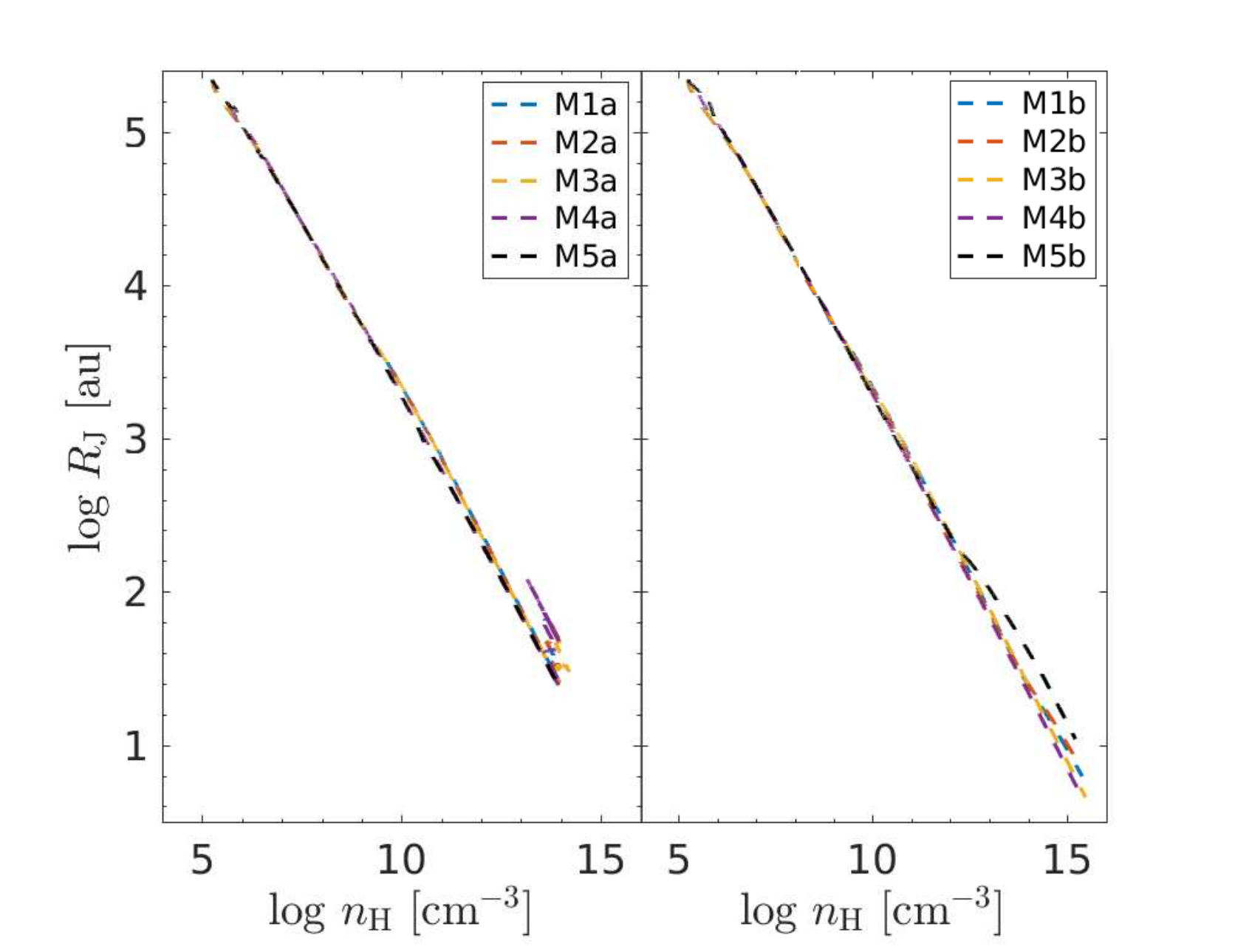}
	\caption{The Jeans length ($R_{\rm J}$) as a function of the gas number density ($n_{\rm H}$) of the collapsing gas cloud. The left and right panels show the results for the models M1a$-$M5a and M1b$-$M5b, respectively. $R_{\rm J}$ and $n_{\rm H}$ are given in units of $\rm au$ and cm$^{-3}$, respectively. }
	\label{fig:figur6}
\end{figure}

\begin{figure}
	\includegraphics[width=\columnwidth]{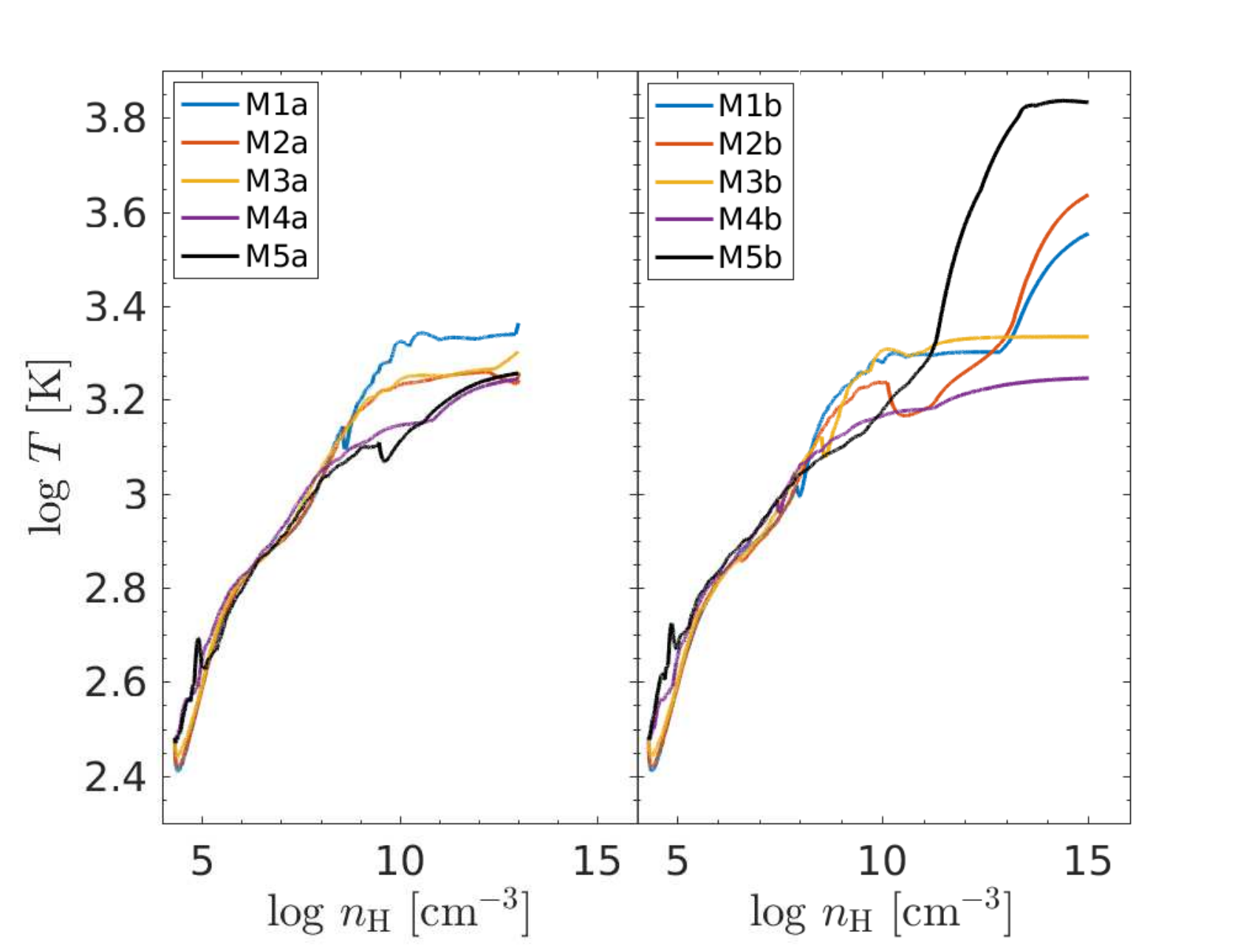}
	\caption{Maximum temperature ($T$) of the collapsing primordial gas cloud as a function of its evolving maximum number density ($n_{\rm H}$). The left and right panels show the results for models M1a$-$M5a and M1b$-$M5b, respectively. The quantities $T$ and $n_{\rm H}$ are provided in logarithmic scale in units of Kelvin and cm$^{-3}$, respectively. }
	\label{fig:figur6}
\end{figure}

\begin{figure}
	\includegraphics[width=\columnwidth]{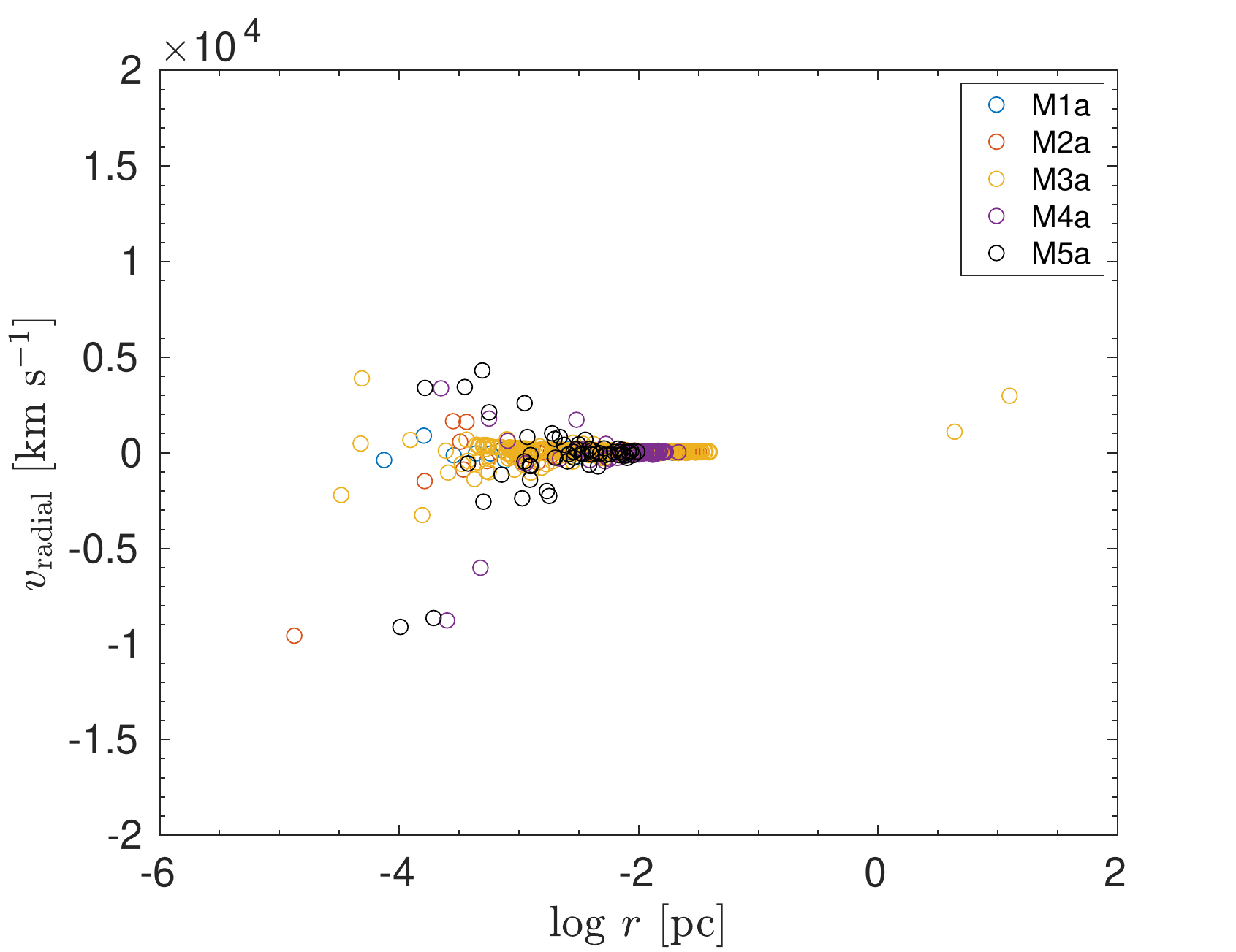}
	\includegraphics[width=\columnwidth]{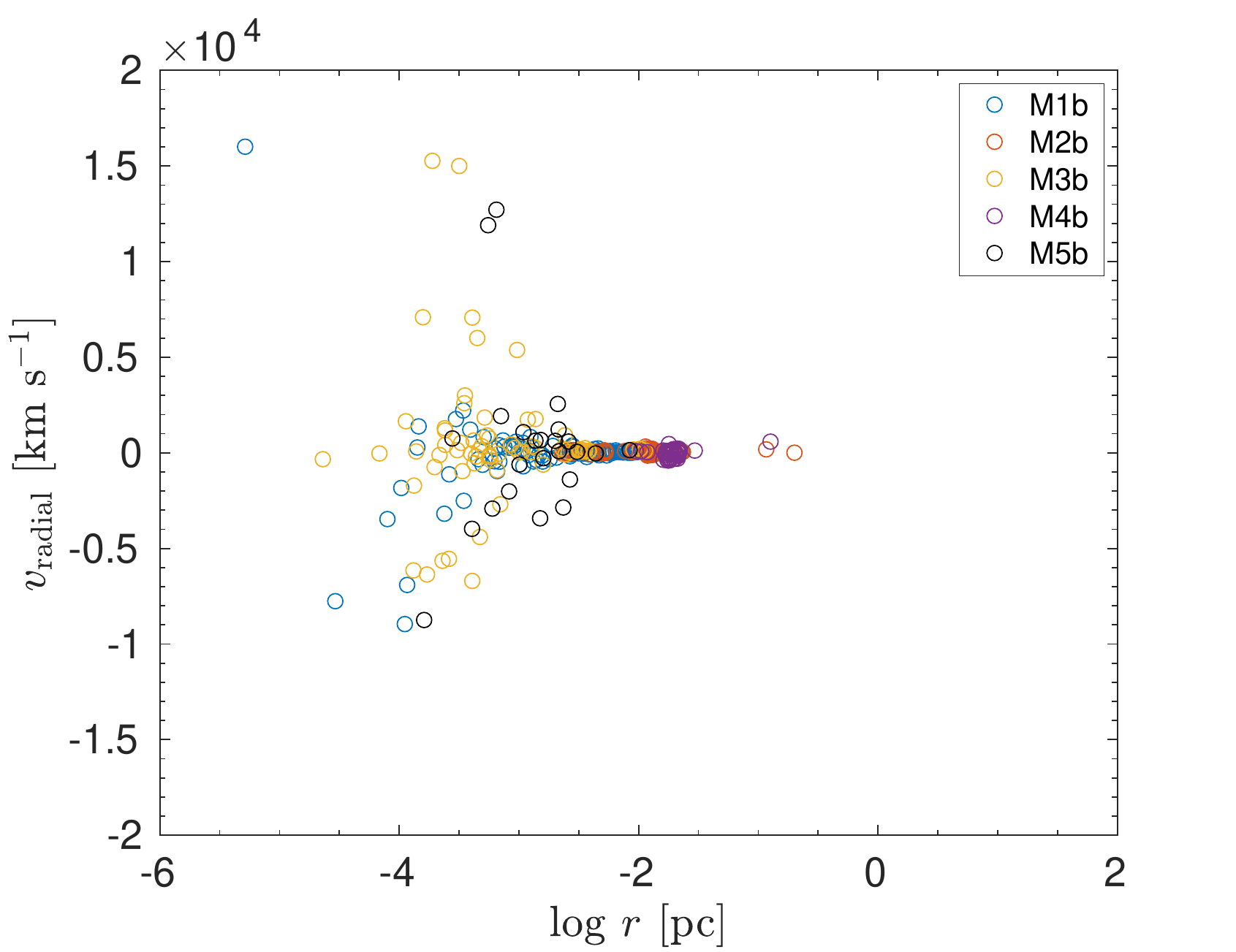}
	\caption{Radial velocity ($v_{\rm rad}$) of the protostars measured from the centre of mass in a cluster of models M1a$-$M5a (top) and M1b$-$M5b (bottom) as a function of the cluster radius ($r$) of the protostars at the time when the SFE reaches  $\xi$ = 2 per cent. The quantity $v_{\rm rad}$ is given in km s$^{-1}$ and $r$ in units of pc. }
	\label{fig:figur6}
\end{figure}

\begin{figure}
	\includegraphics[width=\columnwidth]{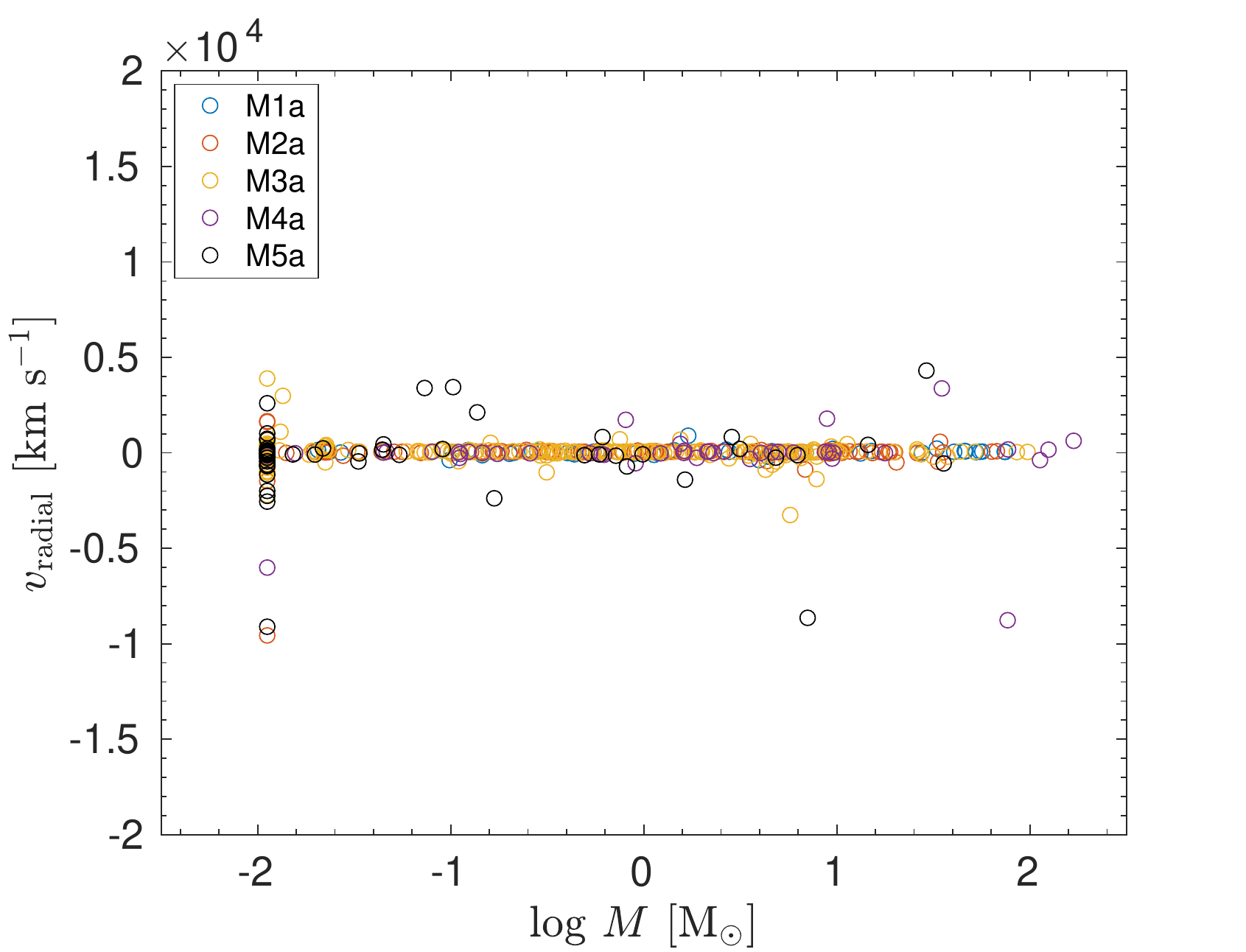}
	\includegraphics[width=\columnwidth]{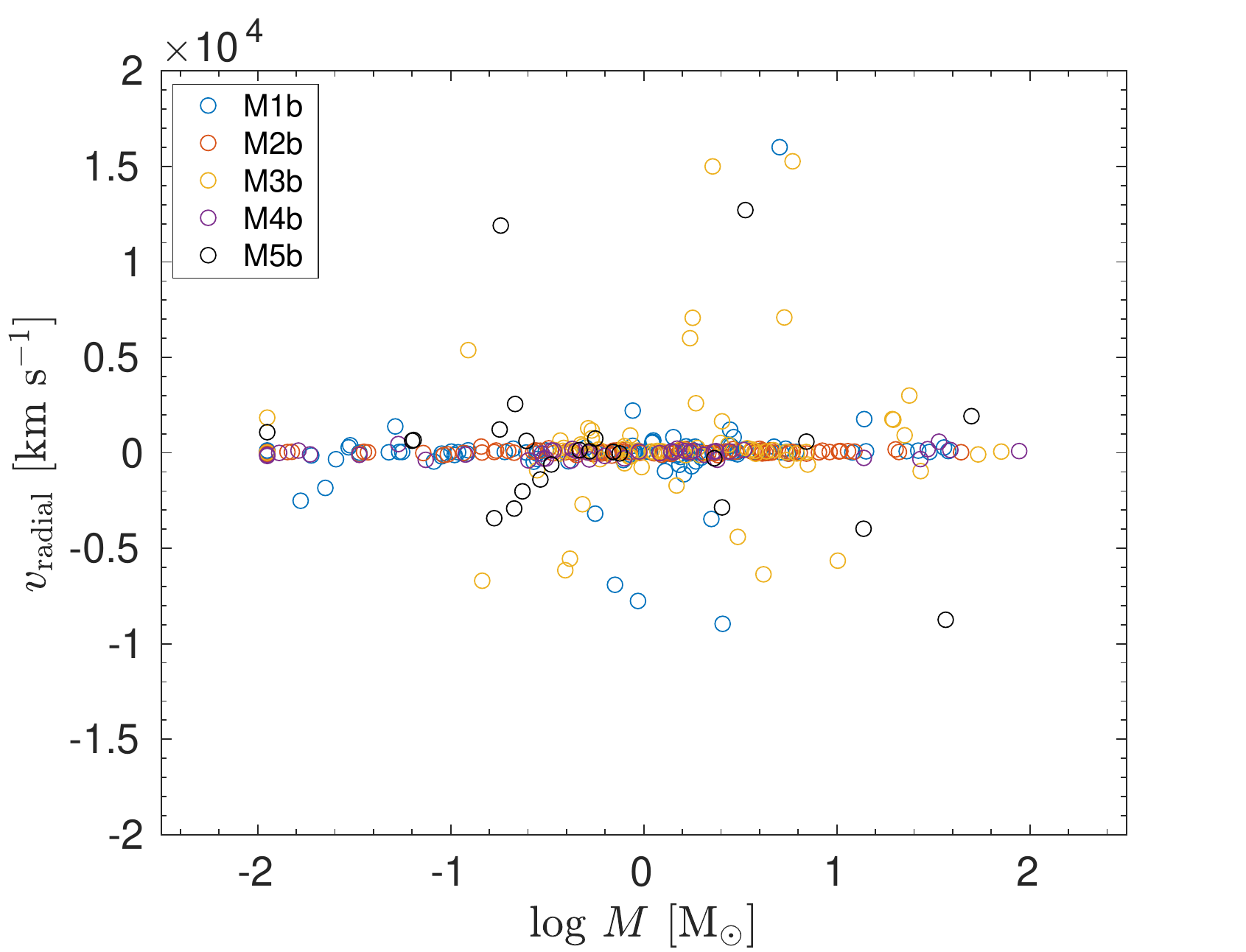}
	\caption{Radial velocity ($v_{\rm rad}$) of the protostars measured from the centre of mass in a cluster of models M1a$-$M5a (top) and M1b$-$M5b (bottom) as a function of mass ($M$) of the protostars at the time when the SFE reaches $\xi$ = 2 per cent. The quantity $v_{\rm rad}$ is given in km s$^{-1}$ and $M$ in units of solar mass M$_{\odot}$. }
	\label{fig:figur6}
\end{figure}

\begin{table}
	\centering
	\caption{Summary of the mass accretion for the two sets of models, M1a$-$M5a and M1b$-$M5b. The table is constructed for MMPBs at the time when $\xi$ = 2 per cent. The columns denote the binary component that is considered (primary/secondary),  the minimum mass accretion rate ($\dot M_{\rm min}$), the maximum mass accretion rate ($\dot M_{\rm max}$), and the mean mass accretion rate ($\dot M_{\rm mean}$) that was obtained during the simulation.}
	\label{tab:Table4}
	\begin{tabular}{cccccc} 
		\hline
		\hline
		 Model & component & $\dot M_{\rm min}$ & $\dot M_{\rm max}$  &  $\dot M_{\rm mean}$     \\
		  &  &  (M$_{\odot}$ yr$^{-1}$) & (M$_{\odot}$ yr$^{-1}$) &  (M$_{\odot}$ yr$^{-1}$)    \\
		\hline
			M1a & primary	&	2.5 $\times 10^{-2}$ &	6.6 $\times 10^{-1}$	&	1.0 $\times 10^{-1}$ \\
		    M1a & secondary	&	3.5 $\times 10^{-2}$ &	5.9 $\times 10^{-1}$	&	1.1 $\times 10^{-1}$ \\
			M2a & primary	&	3.4 $\times 10^{-2}$ &	5.4 $\times 10^{-1}$	&	1.1 $\times 10^{-1}$ \\
			M2a & secondary &	2.1 $\times 10^{-2}$ &	4.6 $\times 10^{-1}$	&	8.7 $\times 10^{-2}$ \\
			M3a & primary 	&	9.4 $\times 10^{-4}$ &	2.0 $\times 10^{-1}$	&	4.8 $\times 10^{-2}$ \\
			M3a & secondary &	1.8 $\times 10^{-4}$ &	2.7 $\times 10^{-1}$	&	5.7 $\times 10^{-2}$ \\
		    M4a & primary	&	4.2 $\times 10^{-4}$ &	6.0 $\times 10^{-1}$	&	1.6 $\times 10^{-1}$ \\
			M4a & secondary	&	3.8 $\times 10^{-4}$ &	2.4 $\times 10^{-1}$	&	7.5 $\times 10^{-2}$ \\
			M5a & primary	&	7.8 $\times 10^{-5}$ &	1.9 $\times 10^{-1}$	&	1.8 $\times 10^{-2}$ \\
			M5a & secondary	&	1.6 $\times 10^{-3}$ &	3.1 $\times 10^{-1}$	&	4.8 $\times 10^{-2}$ \\
			\hline
			M1b & primary	&	8.8 $\times 10^{-5}$ &	25.2 $\times 10^{-1}$	&	1.6 $\times 10^{-1}$ \\
		    M1b & secondary	&	7.7 $\times 10^{-6}$ &	18.1 $\times 10^{-1}$	&	4.2 $\times 10^{-2}$ \\
			M2b & primary	&	2.2 $\times 10^{-3}$ &	15.6 $\times 10^{-1}$	&	2.6 $\times 10^{-1}$ \\
			M2b & secondary &	7.5 $\times 10^{-5}$ &	9.3 $\times 10^{-1}$	&	9.0 $\times 10^{-2}$ \\
			M3b & primary	&	7.3 $\times 10^{-4}$ &	45.6 $\times 10^{-1}$	&	6.2 $\times 10^{-1}$ \\
			M3b & secondary &	3.6 $\times 10^{-4}$ &	34.8 $\times 10^{-1}$	&	2.5 $\times 10^{-1}$ \\
		    M4b & primary	&	6.2 $\times 10^{-7}$ &	11.5 $\times 10^{-1}$	&	2.0 $\times 10^{-2}$ \\
			M4b & secondary	&	1.5 $\times 10^{-6}$ &	15.2 $\times 10^{-1}$	&	2.0 $\times 10^{-2}$ \\
			M5b & primary	&	2.6 $\times 10^{-5}$ &	61.4 $\times 10^{-1}$	&	6.5 $\times 10^{-2}$ \\
			M5b & secondary	&	7.0 $\times 10^{-4}$ &	33.2 $\times 10^{-1}$	&	2.2 $\times 10^{-1}$ \\
		\hline
	\end{tabular}
\end{table}

\begin{table*}
	\centering
	\caption{Summary of the results for the two sets of models M1a$-$M5a and M1b$-$M5b. The table is constructed for MMPBs at the time when the SFE ($\xi$) reaches 2 per cent. The columns indicate the binary component that is considered, the mass of the component, its age, the semi-major axis ($a$), the eccentricity ($e$) and the mass ratio ($q$) at the end of the simulation. 
	Note: The protostars are assigned their status as primary or secondary component based on their final masses at the end of the simulation, and not by their time of creation during the cloud collapse. }
	\label{tab:Table4}
	\begin{tabular}{cccccccc} 
		\hline
		\hline
		Model & component type & component mass (M$_{\odot}$) & age (kyr) & $a$ (au) & $e$ &  $q$    \\
		\hline
		M1a	& primary, secondary &	29.05, 14.87 & 3.03, 1.52 	&	210.9		&	0.63		&	 0.51 \\
		M2a	& primary, secondary &	81.77, 29.96 & 4.52, 4.54 	&	681.4		&	0.41	    &	 0.36 \\
		M3a	& primary, secondary &	62.40, 53.10 & 5.21, 5.27	    &	253.2		&	0.18		&	 0.85 \\
		M4a	& primary, secondary &	68.97, 12.99 & 4.48, 2.94     &	86.05		&	0.68		&	 0.18 \\
		M5a	& primary, secondary &	36.18, 30.55 & 4.43, 4.82     &	88.94		&	0.88		&	 0.84 \\
		\hline
		M1b	& primary, secondary &	36.35, 14.03 & 1.81, 1.49	  & 37.09		&	0.46		&	 0.38 \\
		M2b	& primary, secondary &	12.48, 5.77  & 2.44, 0.33     &	290.0		&	0.68	    &	 0.46 \\
		M3b	& primary, secondary &	71.94, 19.86 & 1.23, 1.18     &	37.2		&	0.40		&	 0.27 \\
		M4b	& primary, secondary &	31.55, 28.85 & 5.20, 5.20     &	3.57		&	0.16		&	 0.91 \\
		M5b	& primary, secondary &	50.41, 13.92 & 2.20, 1.51     &	144.9		&	0.63		&	 0.27 \\
		\hline
	\end{tabular}
\end{table*}

To avoid the “Courant catastrophe",
we introduce sink particles inside the collapsing gas cloud once a certain density threshold is reached \citep{bate1995modelling, bromm2004accretion, federrath2010modeling, stacy2010first}, as outlined in section~\ref{methods}. As mentioned in the introduction, in many of the previous works sink particles have been introduced at gas densities that correspond to the gas phase where H$_{2}$ line cooling still operates  \citep{stacy2010first, clark2011gravitational, dutta2016effects, riaz2018formation, sharda2019role}. The collapsing primordial gas is most susceptible to fragmentation at densities below $ 10^{14}$~cm$^{-3}$ \citep{hartwig2015new}. However, more recently, \citet{wollenberg2020formation} have performed a statistical analysis of an ensemble of simulations of Pop. III star formation where the gas density to form sinks corresponds to the CIE cooling phase. Here we take into account both the H$_{2}$ line cooling phase and the CIE cooling phase. It is worth mentioning that we do not include magnetic fields, nor radiation feedback nor a UV background in our calculations. This will be discussed in more detail in the caveat section. For the comparison runs where the gas only experiences  the H$_{2}$ line cooling (models M1a$-$M5a), we set the sink formation density to $n_{\rm sink}$ $=$ 10$^{13}$ cm$^{-3}$. 

On the other hand, the models that include both the H$_{2}$ line cooling phase and also the subsequent CIE cooling phase (a case that remains relevant in our models M1b$-$M5b), we set $n_{\rm sink}$ $=$ 10$^{15}$ cm$^{-3}$. For these two values of $n_{\rm sink}$, the accretion radius $r_{\rm acc}$ in our simulations is set to a unique value, which for models set M1a$-$M5a and M1b$-$M5b remains 28 au and 5 au, respectively. It is important to keep in mind the possible effect of a larger $r_{\rm acc}$, as it may influence the fragmentation behavior \citep{hartwig2015new}.  

For the chemical composition of the primordial gas we use a total of nine chemical species (H, H$^{+}$, He, He$^{+}$, He$^{++}$, e$^{-}$, H$_{2}$, H$_{2}^{+}$, H$^{-}$) with the respective initial fractional abundances as $f_\mathrm{H}=0.75$, $f_\mathrm{He} = 0.24899$, $f_\mathrm{H^+} = 8.2\times10^{-7}$,  $f_\mathrm{e} = 4.4 \times 10^{-10}$, and $f_\mathrm{H_2} = 10^{-3}$ (see also RBVS). 

\begin{figure}
    \includegraphics[width=\columnwidth]{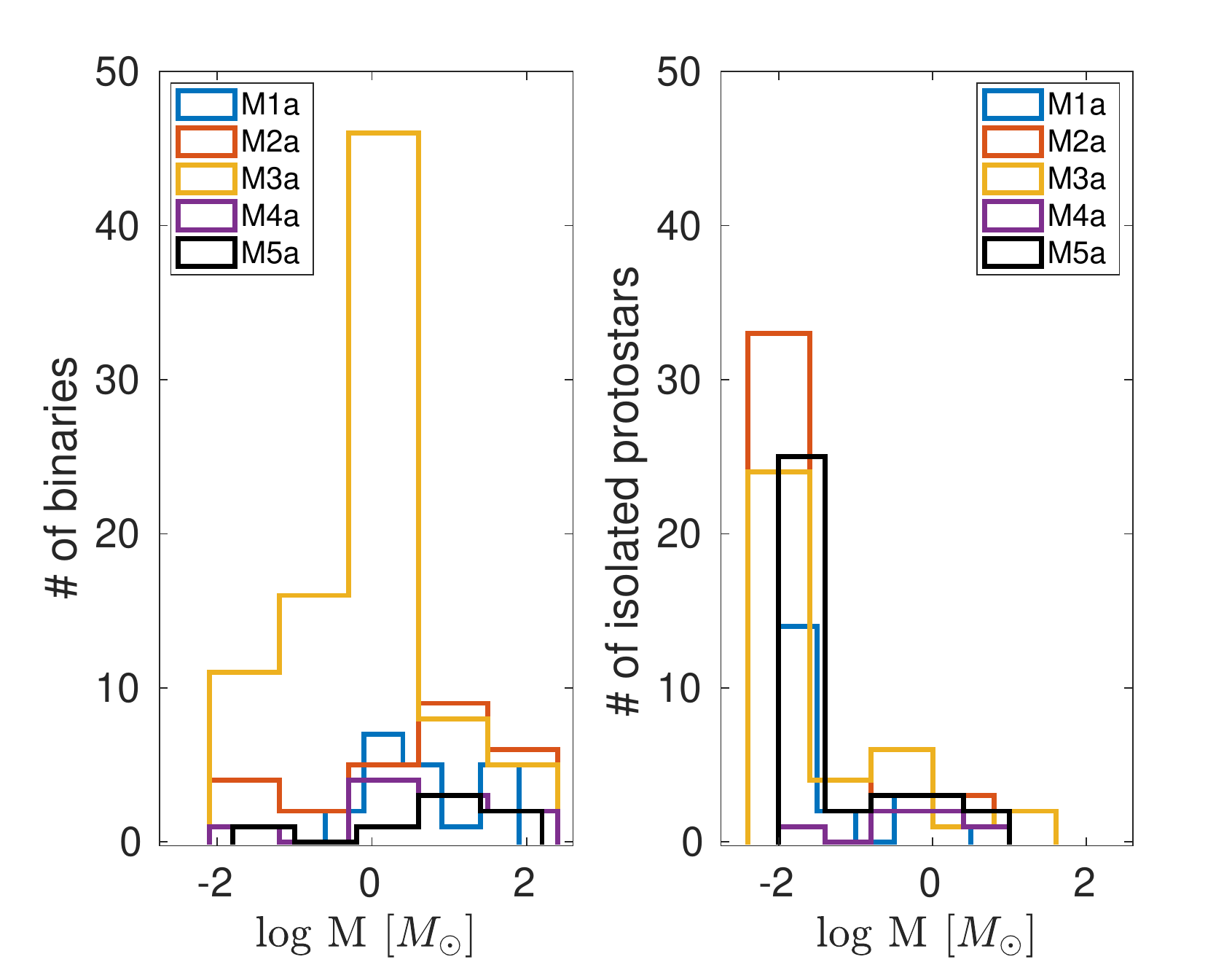}
	\includegraphics[width=\columnwidth]{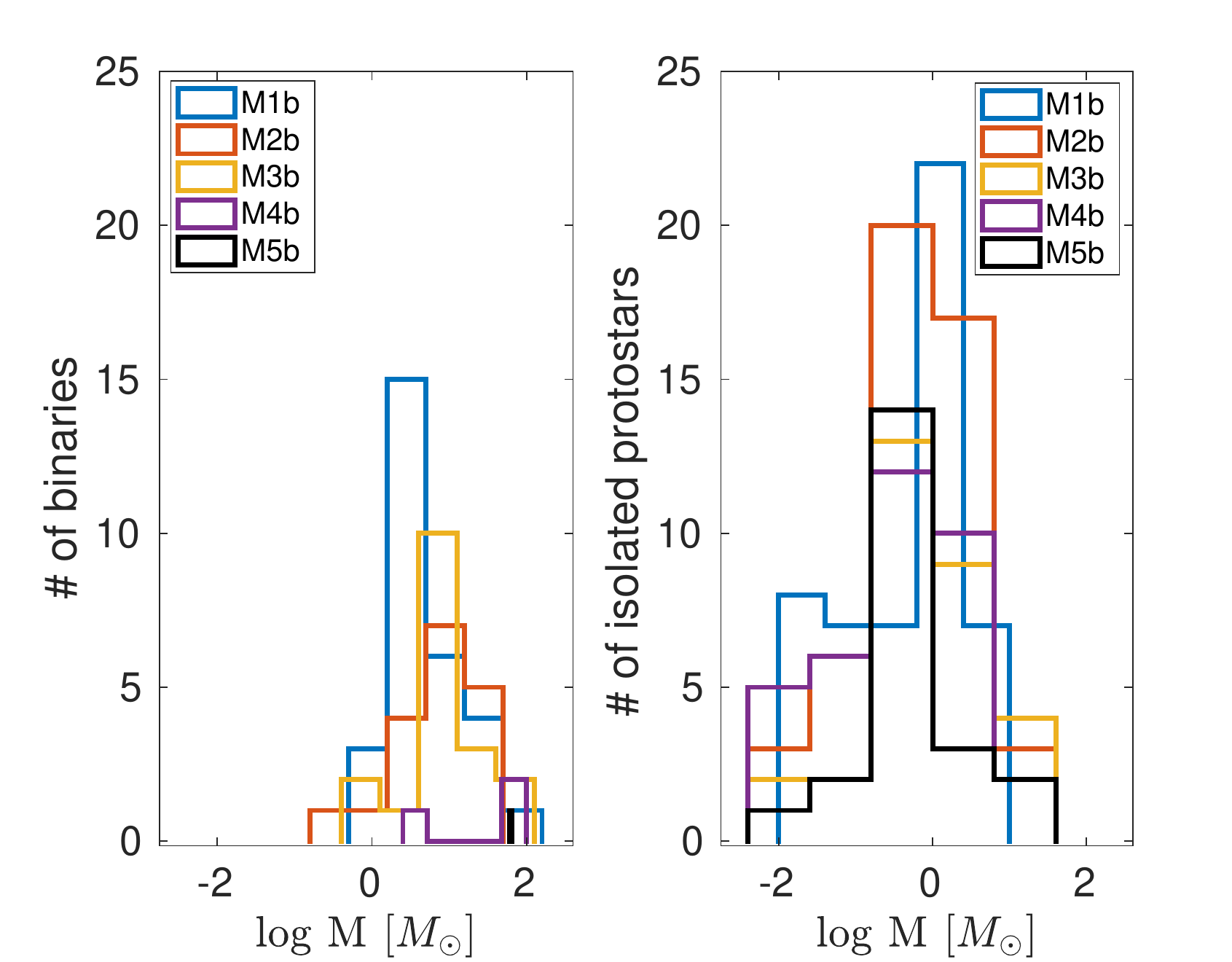}	
	\caption{Top left and right, for models M1a$-$M5a, the number of binary protostars and isolated protostars as a function of their mass ($M$), respectively. Bottom left and right, for models M1b$-$M5b, the number of binary protostars and isolated protostars as a function of their mass ($M$), respectively. The mass of the binary and the isolated protostars is given in units of solar mass M$_{\odot}$. These mass distributions represent the evolutionary stage when  $\xi$ = 2 per cent.}
	\label{fig:figur6}
\end{figure}

\section{Results and discussion}
We now present our simulation results and discuss these in detail. For the visualisation of our simulation results, we use the visualisation tool SPLASH developed by \citet{price2007splash}.

\subsection{Morphology of the collapsing cloud}
Figure 1 maps the density structure at the final stage of our  simulation models M1a$-$M5a and M1b$-$M5b when the SFE reaches $\xi$ = 2 per cent. The arrows in each panel mark the location of the primary and secondary components of the MMPB (i.e. the binary system with the largest total mass based on both companions) in the cluster of protostars in each model. In each set of models, we systematically increase the initial turbulent Mach number from $\mathcal{M}$ = 0.1 to 1. The gravoturbulent collapse of models M1a$-$M5a (top row), i.e. the models which are based only on H$_{2}$ line cooling, leads to the formation of a set of dense gas filaments, which are interconnected and also surrounded by regions of dense gas. A significant number of protostars  leave these filaments due to their dynamical evolution and move to a lower-density part of the cloud. We expect this dynamical evolution of the protostars to also affect their subsequent mass growth. On the other hand, we find the location of the MMPBs in almost each model to still be located inside the dense gas regions of the collapsing gas cloud.   

The panels in the bottom row show the collapse of primordial gas clouds in models M1b$-$M5b that include both cooling phases. A systematic increase in the initial turbulent Mach number from $\mathcal{M}$ = 0.1 $-$ 1 in these models leads to a systematic dispersal of the two spiral-arm structures that eventually become gas filaments when the turbulence in the gas is set to the highest value in our models. The protostars, in general, show scattered positions. The MMPB, however, still remains part of a local high density region in each model. 

Figure 2 shows the thermal structure at the final stage of models M1a$-$M5a and M1b$-$M5b when the SFE reaches $\xi$ = 2 per cent. The locations of the protostars in relatively cold gas regions due to their dynamical evolution  affect their mass accretion process and hence their individual mass growth.   

\subsection{Global disc structure}
We show in Figure 3 the global disc structure emerging for our sets of simulations, i.e. M1a$-$M5a and M1b$-$M5b. For weakly turbulent gas clouds (i.e. $\mathcal{M}$ = 0.1 and 0.2) the gas infall dominates over rotation and the prestellar gas cloud tends to form a disc structure aligned with the rotation axis within which fragmentation can occur \citep{abel2000formation, bromm2001fragmentation, dutta2016effects} (see Figure 3, first two panels in each row). With the increase of the Mach number towards mildly turbulent gas models (i.e. $\mathcal{M}$ = 0.4 $-$ 0.8) the resulting disc structure gradually becomes subject to misalignment with respect to the rotational axis of the gas cloud (see Figure 3, third and fourth panel in each row). Moreover, from mildly subsonic turbulence to transonic turbulence (i.e. $\mathcal{M}$ = 0.8 $-$ 1.0) the disc structure becomes significantly misaligned and is also found to be at the verge of disruption by the time the SFE reaches $\xi$ = 2 per cent (see Figure 3, last two panels in each row).  Thus, the formation of the protostars in weakly turbulent gas clouds is most likely to occur via global disc rotation-induced fragmentation \citep{yoshida2006formation}. In the other models,  gas cooling promotes cloud deformation into sheets or filaments \citep{chiaki2016gravitational}. 

We therefore suggest that the two fragmentation scenarios, namely the disc fragmentation and the filament fragmentation, are closely related to the strength of the turbulence present in the primordial gas cloud. We also emphasize that the global disc we report in our models is a rotating disc, and not an accretion disc. There is no central object present in the disc but a cluster of dynamically evolving protostars.              

\subsection{Thermal structure and Jeans length}
Figure 4 illustrates the computed Jeans length $R_{\rm J}$ from equation 7 for the evolution of the collapsing primordial gas cloud until the formation of the first protostar in each model. The final value of $R_{\rm J}$ in our simulations also serves as an indicator for the scales that are still well-resolved  and hence provides justification for our selection of $r_{\rm acc}$ = 28 au and 5 au in the respective set of models M1a$-$M5a and M1b$-$M5b. $R_{\rm J}$ shows a negative slope as the primordial gas cloud collapses to higher densities and heats up adiabatically.

In Figure 5, we show the relationship between the gas density and temperature for each set of models until the first protostar appears in the collapsing gas cloud. The left and right panels exhibit the temperature as a function of the gas number density $n_{\rm H}$ for models M1a$-$M5a and M1b$-$M5b, respectively. Regardless of the type of cooling (i.e. the pure H$_{2}$ line cooling in models M1a$-$M5a and the additional CIE cooling in models M1b$-$M5b), we notice that the gas attains relatively higher temperatures in cases of  weakly turbulent initial states of the primordial gas cloud compared to the strongly turbulent models, which could be a result of more efficient collapse and faster compression. Moreover, at the verge of the formation of the first protostar the difference in the final temperatures of the collapsing gas clouds is more significant in the models which include both types of cooling as compared to the models where only the H$_{2}$ line cooling is considered. This has consequences for the gas accretion process in the protostars (see equation 8). A rapid mass growth mainly due to gas accretion is still possible even if there are less frequent merger events, in the presence of smaller numbers of fragments. This scenario is evident in models M1b$-$M5b where we find a lower number of fragments $N_{\rm max}$ and hence lower chances of merger events (see also Table 2).          

\subsection{Dynamical evolution of the protostars}
In Figure 6, we show the radial velocity of Pop. III protostars as a function of the radius measured from the centre of mass of the star cluster in models M1a$-$M5a and M1b$-$M5b in the top and bottom panels, respectively. The plots are constructed when the SFE reaches $\xi$ = 2 per cent. In both sets of models, the majority of the Pop. III protostars are bound and hence exhibit nearly zero radial velocities. In the top panel in models M1a$-$M5a, a few protostars show higher negative than positive radial velocities and escapers seem to be disfavored. Turbulence in these models seems to play no significant role in defining the resulting radial velocity magnitudes at $\xi$ = 2 per cent  as no difference is found in the radial velocities of the protostars when comparing them for different levels of initial turbulence.

In the bottom panel in models M1b$-$M5b, a relatively larger number of protostars shows non-zero radial velocity magnitudes. However, the turbulence in these models plays hardly any role in the resulting radial velocity magnitudes at $\xi$ = 2 per cent,  as again no difference can be found in the radial velocities of the protostars when comparing models with different levels of the initial turbulence. However, it is interesting to note that despite the strong gravitational potential in models M1b$-$M5b, a few of the protostars exhibit higher positive velocities, which is indicative of escaping Pop. III protostars from the protostellar cluster. In general, with the exception of a few escapers, we find that the Pop. III protostars formed in the H$_{2}$ line cooling regime acquire radial velocities approximately in the range of -10 km s$^{-1}$ $\leq$~$v_{\rm rad}$~$\leq$ 5 km s$^{-1}$ whereas in the models with CIE cooling, the range approximately remains -10 km s$^{-1}$ $\leq$~$v_{\rm rad}$~$\leq$ 20 km s$^{-1}$).

Figure 7 shows the radial velocities of Pop. III protostars (measured from the centre of mass of the star cluster) as a function of their masses. The top and bottom panels illustrate the results of models M1a$-$M5a and M1b$-$M5b, respectively, when the SFE reaches $\xi$ = 2 per cent. A comparison of the two sets of models, in general, reveals no influence of the initial turbulence and its effect on the resulting radial velocities as a function of their mass spectrum. In both sets of models, the mass range is roughly 0.01 $-$ 100 M$_{\odot}$. However, taking into account the impact of both types of gas cooling, we find that in models M1b$-$M5b with both H$_{2}$ line cooling and CIE cooling, some protostars in particular in the intermediate mass range show significant variations in their radial velocities, an effect that seems strongly suppressed for the models  M1a$-$M5a with only H$_{2}$ line cooling. The majority of the protostars in both sets of models exhibits a close-to-zero radial velocity magnitude, indicating a gravitationally bound state of the cluster. The significant positive radial velocities found in some cases have implications for the mass-growth of these protostars. They will stop accreting when leaving the gaseous part of the cluster and may subsequently enter the main sequence as low-mass Pop. III stars \citep{greif2011simulations}. 

 With the dynamical three-body interactions inside the star cluster, less massive protostars can be ejected from the three-body system, leaving behind a compact protobinary system as a result of binary hardening due to the ejected third body, and then eventually from the star cluster itself. However, if a star cluster remains embedded then any protostar ejected from the three-body system will still have to pass through the gas cloud and may continue to accrete until it finally leaves the cloud. We have found in our simulations that in collapsing primordial gas clouds for which both the H$_{2}$ line cooling and the CIE cooling are active, intermediate mass objects acquire higher positive and negative radial velocities. This can influence (delay) the mass segregation in the Pop. III stars cluster.

 In a star cluster that is already shifted away from the gaseous part of the system (hence it is no longer embedded, see Figure 1), any ejected protostar(s) via three-body interaction can immediately stop accreting from the environment as it will not be passing through the gaseous surroundings. We therefore suspect that modeling the collapse of primordial clouds with only H$_{2}$ line cooling can overestimate the number of ejected Pop. III protostars. Previous work suggested that they can still be surviving today, primarily due to their plausible low-masses at the time of their ejection from the parent star cluster \citep{suda2006nucleosynthetic, ishiyama2016low, magg2018predicting, magg2019observational, kirihara2020cosmological}. More recently in a magnetohydrodynamical simulation, \citet{sharda2020importance} and \citet{2022arXiv220102225S} argued that magnetic fields can suppress the formation of low-mass Pop. III stars, thus resulting in a top-heavy Pop III IMF and explaining the absence of observed Pop. III stars at redshift z $=$ 0. However, we believe that more realisations are required to statistically quantify the connection between the turbulence in the primordial gas clouds and the fate of embedded Pop. III star cluster.   

\begin{figure}
	\includegraphics[width=\columnwidth]{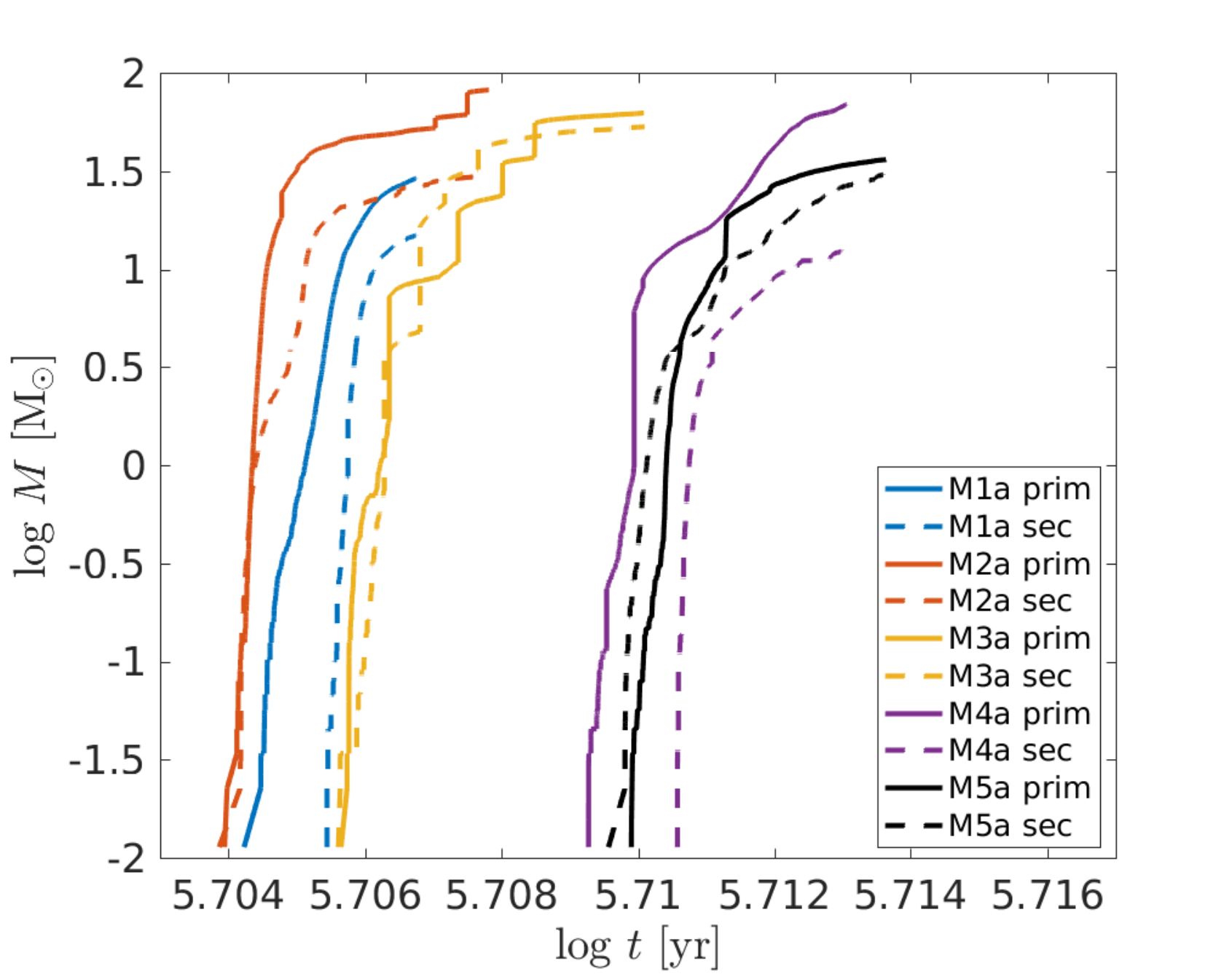}
	\includegraphics[width=\columnwidth]{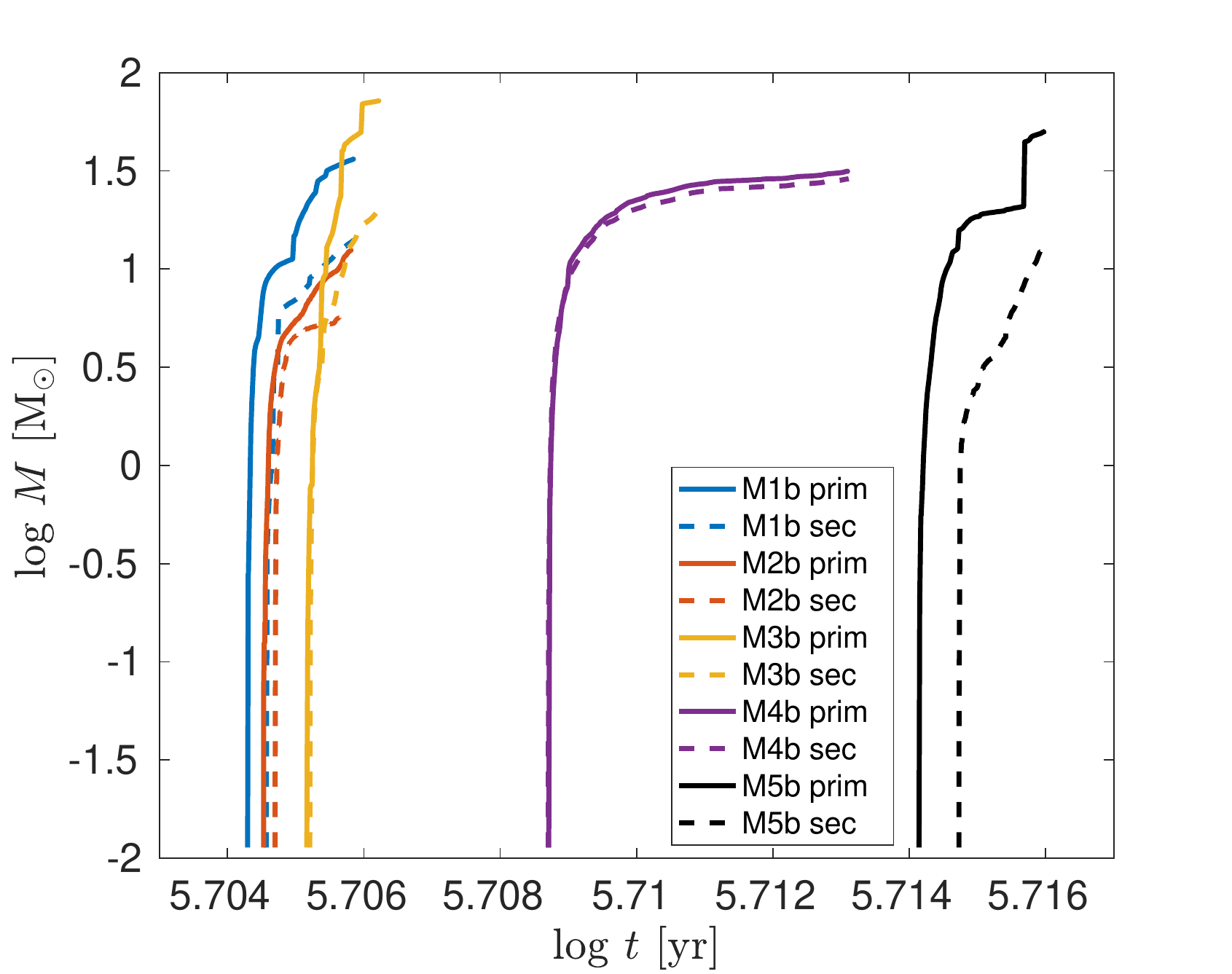}
	\caption{The mass accumulation history of the binary companions that constitute the MMPB in models M1a$-$M5a (top panel) and M1b$-$M5b (bottom panel) at the end of the simulations. The total gas mass converted into the mass of the companions is given in units of solar mass M$_{\odot}$ and the time $t$ in units of yr, in logarithmic scales. }
	\label{fig:figur6}
\end{figure}

\begin{figure}
	\includegraphics[width=\columnwidth]{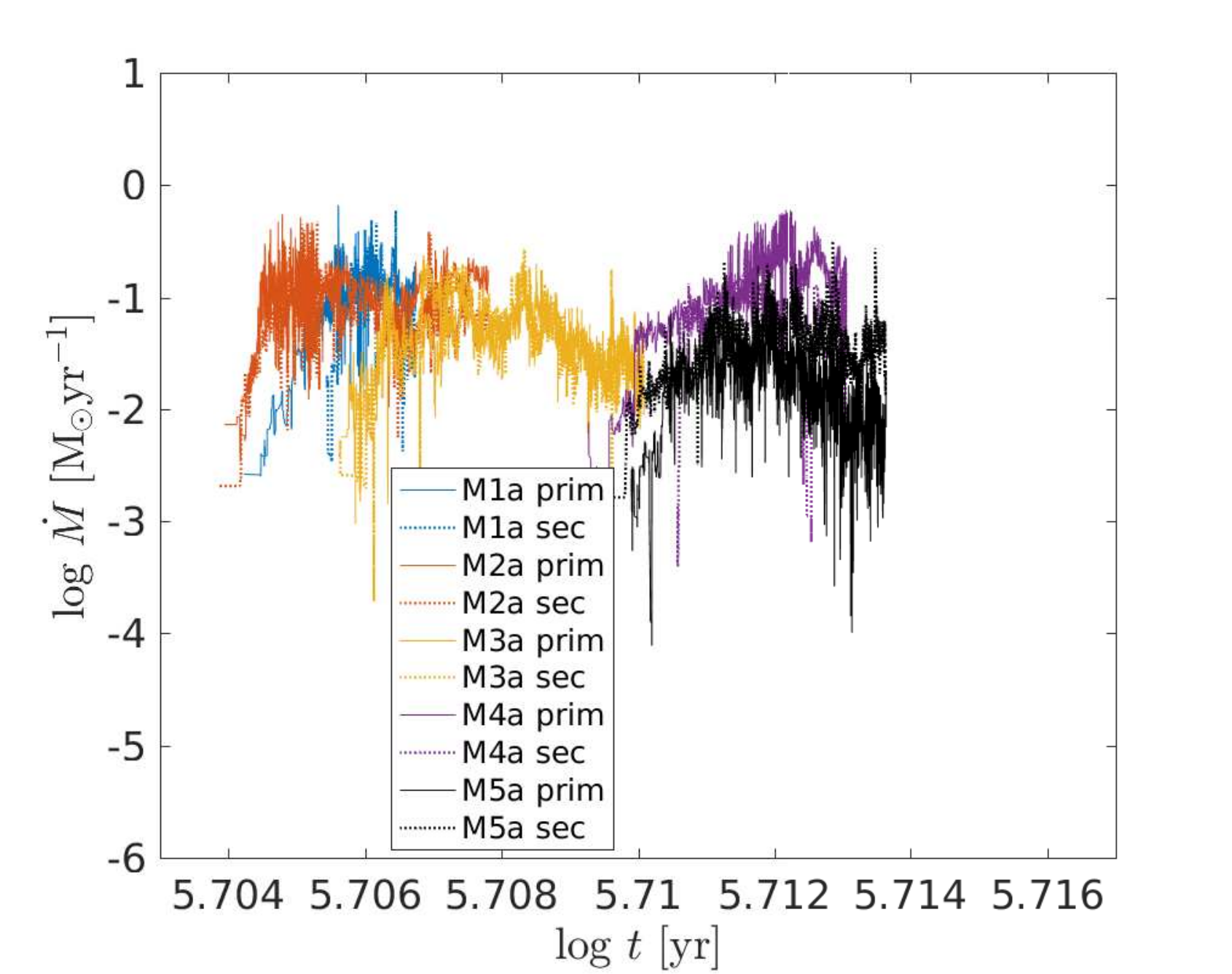}
	\includegraphics[width=\columnwidth]{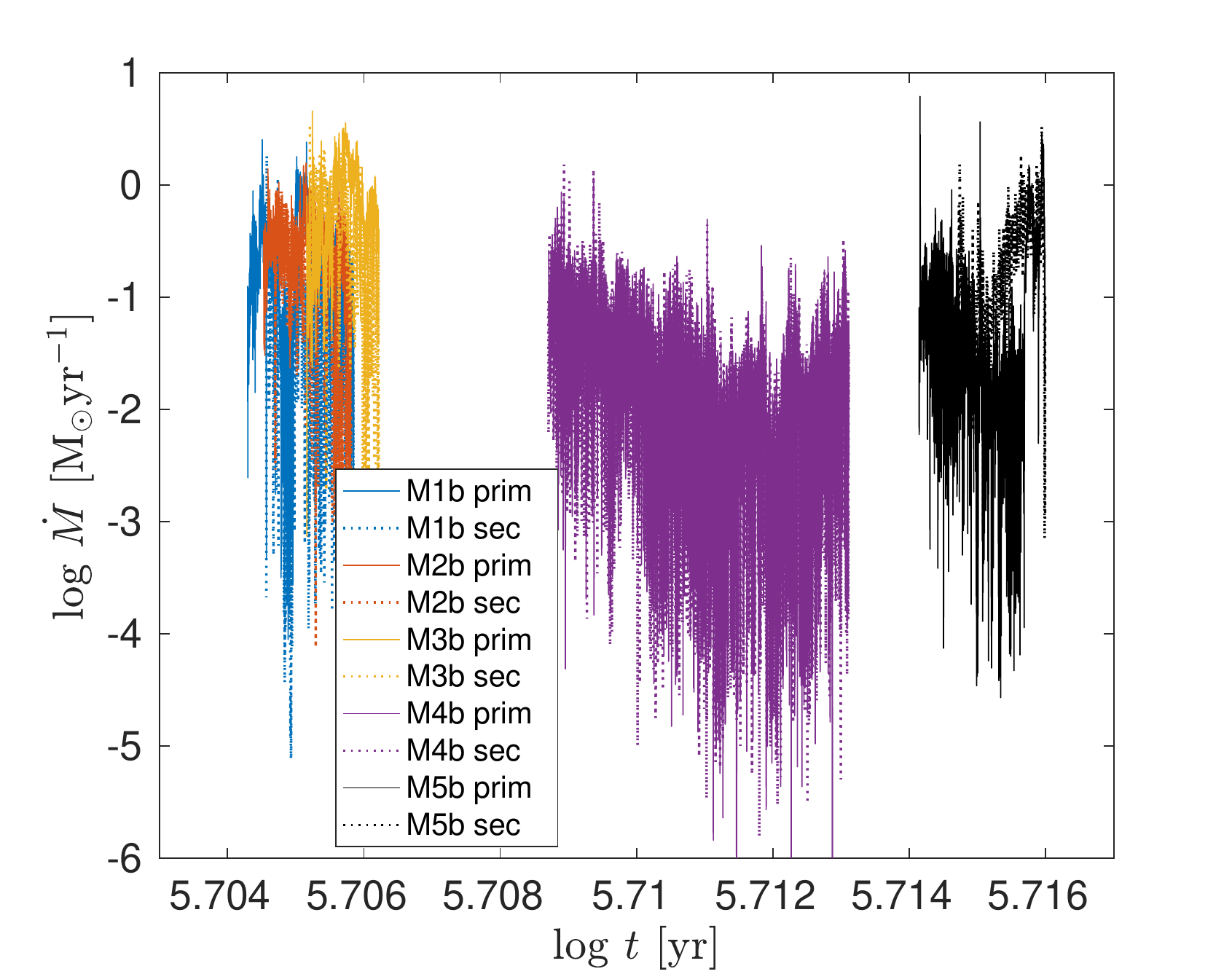}
	\caption{For the MMPBs, the accretion rates $\dot M$ of the primary and secondary companions in models M1a$-$M5a (top) and in models M1b$-$M5b (bottom). The quantities $\dot M$ and $t$ are given in logarithmic units of M$_{\odot}$ yr$^{-1}$ and year, respectively. }
	\label{fig:figur6}
\end{figure}

\subsection{Turbulence, cooling, and the star count}
Protostars can form in large already grown massive Pop. III star associations. This can be a consequence of Jeans unstable clumps still present in the collapsing primordial gas. These clumps can migrate towards the deep potential well of the massive stars \citep{inayoshi2014does, latif2015disc}. In addition to this, fragmentation can also occur in the massive protostellar disc \citep{clark2011formation, ishiyama2016low}. In any case, the massive protostars in these associations have the potential
to disrupt the remaining star-forming gas \citep{inayoshi2014does}.
In a cluster environment, protostars may have possible chaotic orbital motions. Also, via dynamical interactions, the ejected lower-mass protostars from the triplets or higher-order protostellar orbital configurations can add linear motions of fast-moving (ejected) protostars in random directions within the cluster. These non-linear dynamical features have consequences for the protostellar mergers and introduce a fair degree of uncertainty towards the dynamics of the system. The final number of protostars $N_{\rm proto}$ in our simulations is a result of a series of merger events taking place during the dynamical evolution of the cluster. A highly chaotic state of motion of the protostars within the dense cluster remains inevitable. This makes it difficult for the gas cloud to exhibit any systematic trend in $N_{\rm proto}$ with respect to the turbulent Mach number of the gas. In the outcomes of our simulations, we find a weak dependence of $N_{\rm proto}$ on $N_{\rm max}$ in both sets of models M1a $-$ M5a and M1b $-$ M5b (see Table 2). However, a more sound statistical analysis is required to further verify this claim. Also, we would need to model a greater number of realisations to support our finding that there is no correlation between the original number of protostars and the final number of protostars after the merging events take place.

We believe that it is important to unveil a connection (if there is any) between the final star count in a newly formed star cluster and the state of the turbulence of the collapsing gas cloud. Such a connection can even shed light on the phenomenon of mass segregation, which essentially is the most likely dynamical aspect that accompanies the evolution of a star cluster \citep{lin1993protostars, bonnell1998mass, allison2009dynamical, olczak2011highly}. The Jeans mass is controlled primarily by the thermal and turbulent characteristics of the prestellar cloud \citep{li2004formation, pan2009temperature}. This can also affect the mass segregation in the cluster. Frequent merger events that by default support the formation of massive self-gravitating objects can quickly form a dense nucleus constituted of massive protostars. Contrary to this, a collapsing gas cloud that allows fragmentation at lower Jeans masses requires more merger events to form massive protostars if gas accretion is inefficient. This translates into a larger dynamical time that is required before the cluster forms a group of massive fragments, which after following dynamical mass segregation constitutes a nucleus comprised of massive protostars. This has a direct influence on our understanding of the mass segregation process taking place in primordial environments when H$_{2}$ line cooling and CIE cooling are being considered. The former, in general, has the potential to require more dynamical time to form a nucleus of primordial stars than the latter, primarily due the presence of a greater number of both isolated and binary low-mass objects even when the primordial gas clouds have reached $\xi$ = 2 per cent. We emphasize the dynamical evolution of the cluster because the strong stellar feedback from the inner part of the cluster where massive objects that reside can cause expulsion of the primordial residual gas that still surrounds the young cluster at the early stages of its life \citep{matzner2002role, keto2007formation, peters2010h, klassen2012simulating, silich2018gas}. Also, the mass loss due to stellar evolution can have a significant impact on both the structure and survival chances of such a cluster \citep{hills1980effect,boily2003impact, vesperini2010star}.

There is an agreement between observations and theory that most stars are born in small-order multiple systems where the fragmentation of filaments, dense cores and massive accretion discs is the primary origin of multiplicity \citep{offner2022origin}. However, due to  the scope of this paper, we focus mainly on the protobinary systems. Binary systems play an important role concerning the mass loss that occurs in these clusters \citep{van2011mass}. These systems can be of great significance since they provide a channel for the production of compact binaries via the mass-loss-induced eccentric Kozai mechanism \citep{shappee2013mass}. Such compact stellar systems may then act as progenitors of gravitational waves sources.  
In our simulations, the primordial gas cooling under the influence of H$_{2}$ and also in the CIE cooling phase shows no connection between the binary fraction and the cooling mechanism in play. It seems that the number of binaries that appears during the gravitational collapse and hence their contribution towards the SFE neither appears connected to the level of the initial subsonic turbulence  nor to the type of cooling that controls the thermodynamics of the collapsing primordial gas cloud (see Table 2). We, however, would need more model realisations to verify this result. 

\subsection{Mass distributions}
We now discuss the mass distribution of both isolated and binary protostellar configurations as shown in Figure 8 in detail. We present the total mass of the Pop. III protostars in both configurations (binary and isolated) at the end of our simulations. Figure 8 in the top and bottom panels shows the protostellar mass associated with the binary (left-panel) and isolated (right-panel) configurations in models M1a$-$M5a and M1b$-$M5b, respectively. We first compare the top-left and bottom-left panels where the mass going into the binary configuration along with the number of binaries is shown for the models evolving under only H$_{2}$ line cooling and the ones including also the  CIE cooling. We find that the mass going into the binaries can be an order of magnitude smaller in the pure H$_{2}$ line cooling models compared to the simulations that also include the CIE cooling mechanism. 
 
In the context of the turbulence, we do not see any systematic trend for the number of protobinary systems in models M1a$-$M5a where the cooling is solely provided by the H$_{2}$ line radiation. However, in models M1b$-$M5b we see a hint of systematic trend for the number of protobinary systems such that the binary number decreases with increasing strength of the turbulence in the primordial gas cloud. Moreover, at the time when SFE reaches $\xi$ = 2 per cent, our models that follow only the H$_{2}$ line cooling cover a wider mass range compared to the models that in addition to the H$_{2}$ line cooling also take into account the CIE cooling. A similar comparison but for the isolated Pop. III protostellar configuration is shown in the top-right and bottom-right panels of the same figure. The number of isolated protostars with smaller masses is greater in the gas collapsing under sole H$_{2}$ line cooling compared to the case of CIE cooling. Also in the former, the number of isolated protostars does not follow a systematic trend with respect to the initial state of turbulence. Similarly, models which take into account both the H$_{2}$ line cooling followed by cooling via the CIE do not clearly show any trend. However, the peak in the number of isolated protostars remains close to a solar mass (1 M$_{\odot}$). Also, weakly turbulent gas models yield a greater number of isolated protostars than highly turbulent gas models.

\subsection{Mass accumulation in binary components}
We now focus on the binary systems that appear in our models at the end of the simulations and discuss the mass accumulation for the binary companions. First, we select the MMPB that forms in each model. We then classify its binary components according to their final masses and mark them as primary and secondary companions (see Table 3). At the time of Pop. III star formation, these could have masses as small as $10^{-2}$ M$_{\odot}$ \citep{palla1983primordial, omukai1998formation}.  Figure 9 indicates the time evolution of the binary components as they continue to accumulate mass from the surrounding gas until the simulations are terminated (i.e. at $\xi$ = 2 per cent). The top and the bottom panels show the mass evolution of the two components of the MMPBs that exist in model sets M1a $-$ M5a and M1b $-$ M5b, respectively. The sudden bumps  in the mass accumulation history of the binary components are indicative of  protostellar merger events and otherwise the smooth increasing trend in the mass of the components indicates gas accretion from the surroundings that helps the components to grow in their masses. In the top panel, the mass accumulation for the two components of the MMPB in each model generally remains a gradual process. The mass growth of the MMPBs for the weakly turbulent models exhibits a systematic trend. However, for strongly turbulent clouds the components and hence the binary system itself requires longer times to accumulate mass in the collapsing primordial gas cloud as we increase the strength of the turbulence in these models. In the bottom panel, we observe a systematic trend for the mass accumulation over time for both components of the MMPBs.  We find that the evolving masses of the binary components exhibit a much steeper growth than in the top panel. This indicates efficient gas accretion even in the absence of frequent merger events. For models M1b$-$M5b, the mass growth of the MMPBs exhibits a systematic trend such that the weakly turbulent gas models take less time for one of the binary systems to become very massive. As we increase the strength of the turbulence in the primordial gas it systematically requires more time for a binary in each model to exhibit significant mass growth. In general, Pop. III protobinary systems can become very massive in simulations that include both H$_{2}$ line cooling as well as the subsequent CIE cooling.     
    
\subsection{Accretion rates in binary components}
We quantify the mass accretion rate and its evolution for each component of the MMPB in our models. \citet{shu1977self} provides a useful estimate for the protostellar accretion rate that can be derived from the relation

\begin{equation} \label{transition}
\dot M \backsimeq \frac{M_{\rm J}}{t_{\rm ff} } \backsimeq \frac{c_{\rm s}^{3}}{G} \varpropto T^{3/2}, 
\end{equation}
where $\dot M$ is the mass accretion rate of a protostar,   $M_{\rm J}$ is the Jeans mass, $c_{\rm s}$ is the sound speed, $G$ the gravitational constant, and $T$ the gas temperature. 

It has been well established that the lack of metals and dust in the primordial gas causes Pop. III protostars to accrete at much higher rates than observed in the present-day star-forming gas \citep{nishi1998thermal, mac2004control, mckee2007theory, greif2011simulations}. Figure 10 shows the mass accretion rates of the MMPBs in models M1a $-$ M5a and M1b $-$ M5b in the top and bottom panels, respectively. Generally, in the models that follow both the H$_{2}$ line cooling and subsequent CIE cooling, the binary components exhibit a more vigorous mass accretion compared to MMPBs evolving in the sole H$_{2}$ line cooling models. The strong and rapid mass accretion activity observed in models M1b$-$M5b systematically follows the turbulence in the primordial gas clouds. Weakly turbulent gas clouds enable the massive protobinary systems to accrete fairly rapidly when compared with the MMPBs which evolve in models where only the H$_{2}$ line cooling mechanism dictates the gas collapse (i.e. models M1a $-$ M5a). This phenomenon seems well connected with the overall temperature-density relationship that we presented and discussed in Figure 5.   

In Table 3, we quantify the mass accretion rates resulting from our two sets of models. The minimum mass accretion rate $\dot M_{\rm min}$ for the MMPBs in the sole  H$_{2}$ line cooling models on average remains more than two orders of magnitude larger than for the models that follow both the H$_{2}$ line cooling plus CIE cooling. However, the maximum mass accretion rate $\dot M_{\rm max}$ for the MMPBs in the latter type of models shows more than an order of magnitude higher values as compared to the sole H$_{2}$ line cooling models. For the mean accretion rate $\dot M_{\rm mean}$, we find a weak dependence on the nature of the cooling mechanism. $\dot M_{\rm mean}$ on average fluctuates around 0.81 $\times 10^{-1}$ M$_{\odot}$ yr$^{-1}$ and 1.74 $\times 10^{-1}$ M$_{\odot}$ yr$^{-1}$ in the sole H$_{2}$ line cooling models (M1a$-$M5a) and the H$_{2}$ line cooling plus CIE cooling models (M1b$-$M5b), respectively. This is indicative that the average accretion rate does not significantly depend on the cooling mechanism and the level of turbulence. 

\subsection{Orbital properties}
In Figure 11 we present the properties of the Pop. III protobinary star systems that form in our simulations. The binary systems in our models M1a$-$M5a and M1b$-$M5b cover a range of semi-major axis 3.57 au $\leq$~$a$~$<$ 1000 au. We find no evidence for a systematic variation in the semi-major axis $a$ of all the binary systems with the initial turbulence in the primordial gas cloud. However, the cooling mechanism in the primordial gas can influence the distribution of the semi-major axis as
the peak of the distribution of $a$ shifts more towards the compact binaries in models M1b $-$ M5b compared to models M1a $-$ M5a (see top panels in Figure 11). The MMPBs in our simulations cover a range of semi-major axis 3.57 au $\leq$~$a$~$<$ 681.4 au (see Table 4). It is worth noticing that the minimum resolvable Jeans length which also serves as the constant accretion radius $r_{\rm acc}$ of the protostars in our calculations is 5 au for models M1b $-$ M5b. The MMPB in model M4b with its  $a$ = 3.57 au suffers from the limited spatial resolution. However, the rest of the MMPBs in both sets of models remains fully spatially resolved. Many of the MMPBs in our simulations reside in a dense gas structure. Hence, the dynamical interaction between the binary system and the gaseous medium can induce a braking torque, which leads to the shrinking of the orbits of the binary components due to the extraction of energy and angular momentum from the binary \citep{kim2008dynamical, stahler2010orbital, sanchez2014binaries}. Therefore, in our simulations, the existence of the MMPB inside the dense gas structure is indicative of a possible orbital decay over time, mainly due to gas dynamical friction that may eventually transform it into a hard Pop. III binary. 

\begin{figure}
	
	\includegraphics[width=\columnwidth]{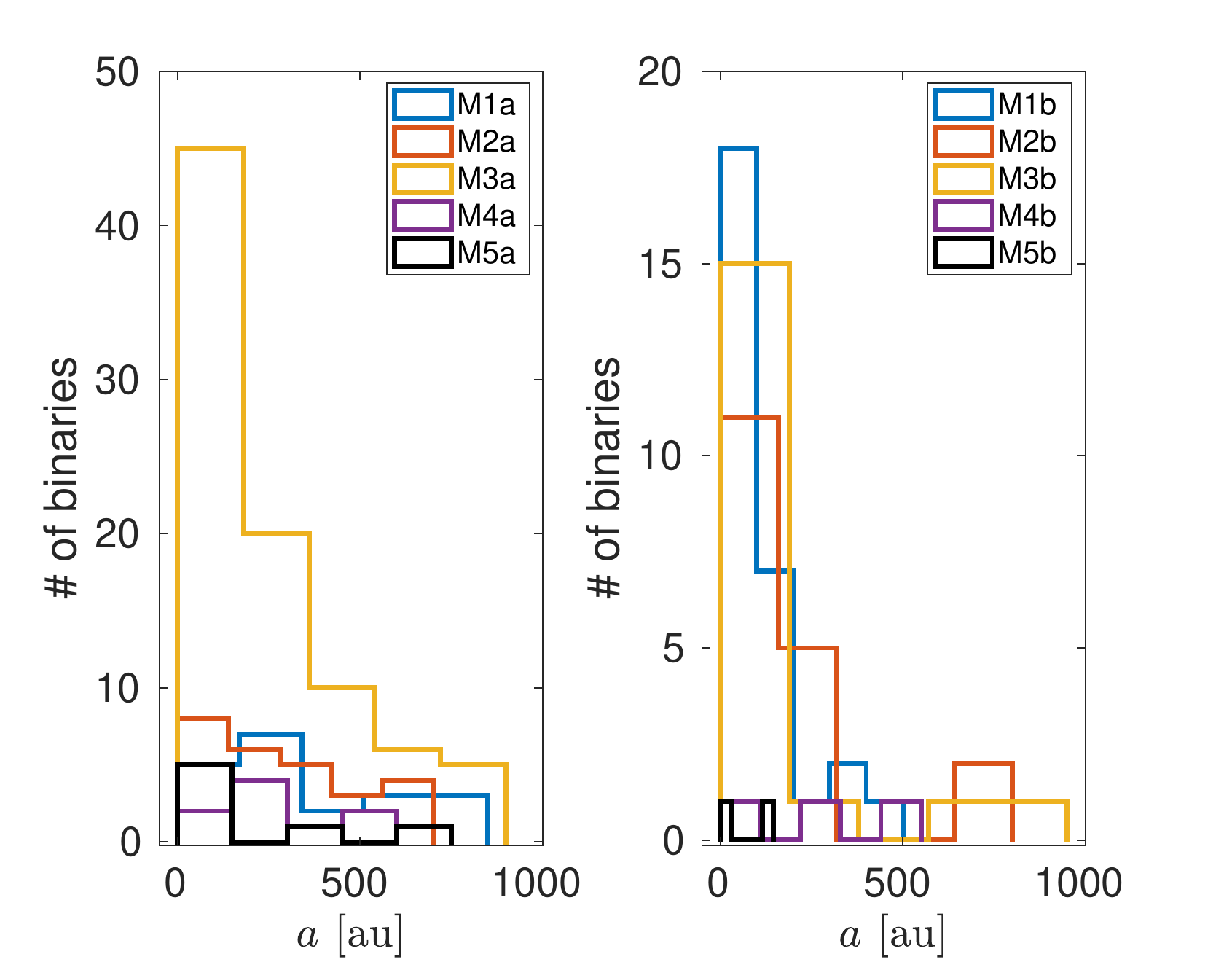}
	\includegraphics[width=\columnwidth]{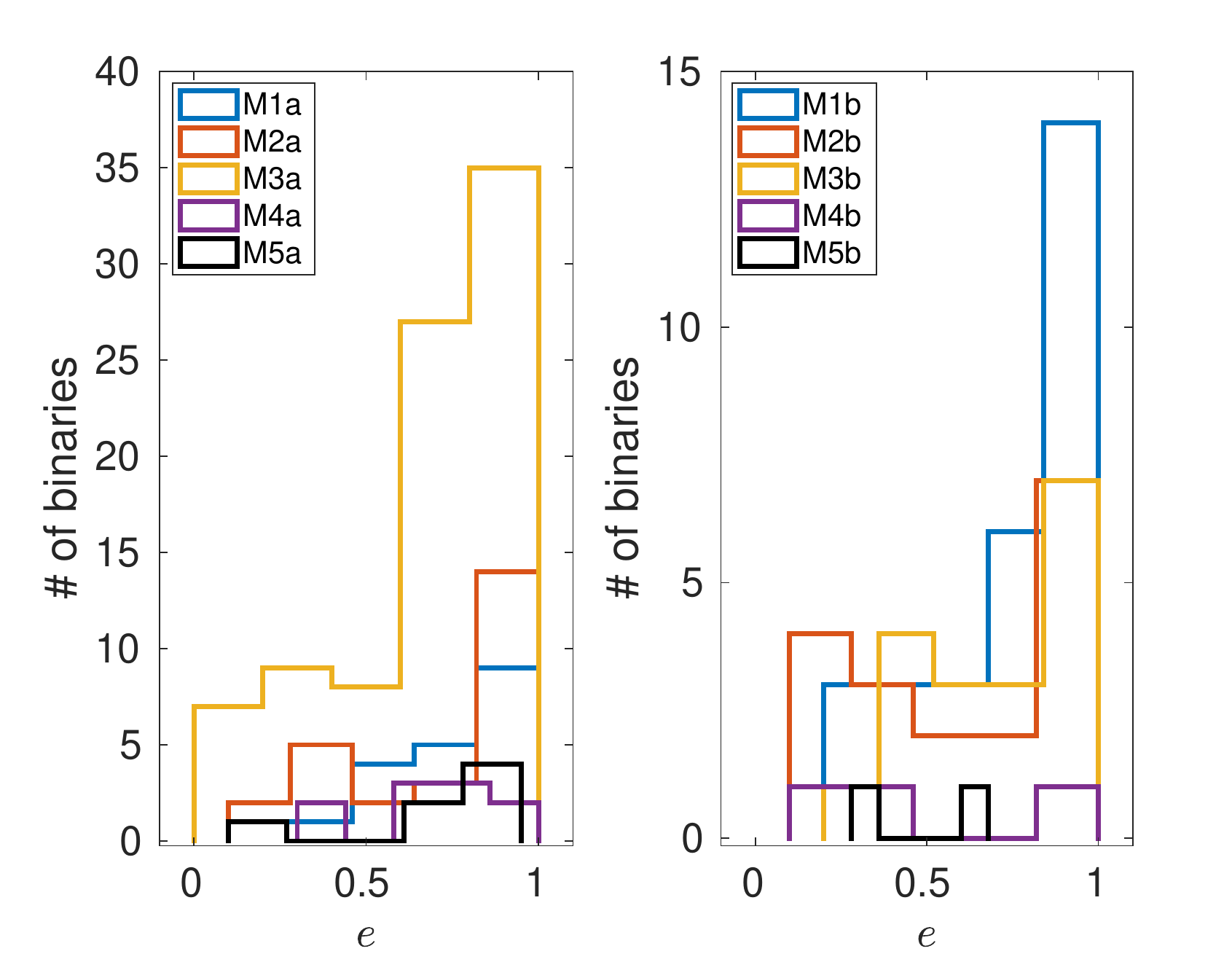}
	\includegraphics[width=\columnwidth]{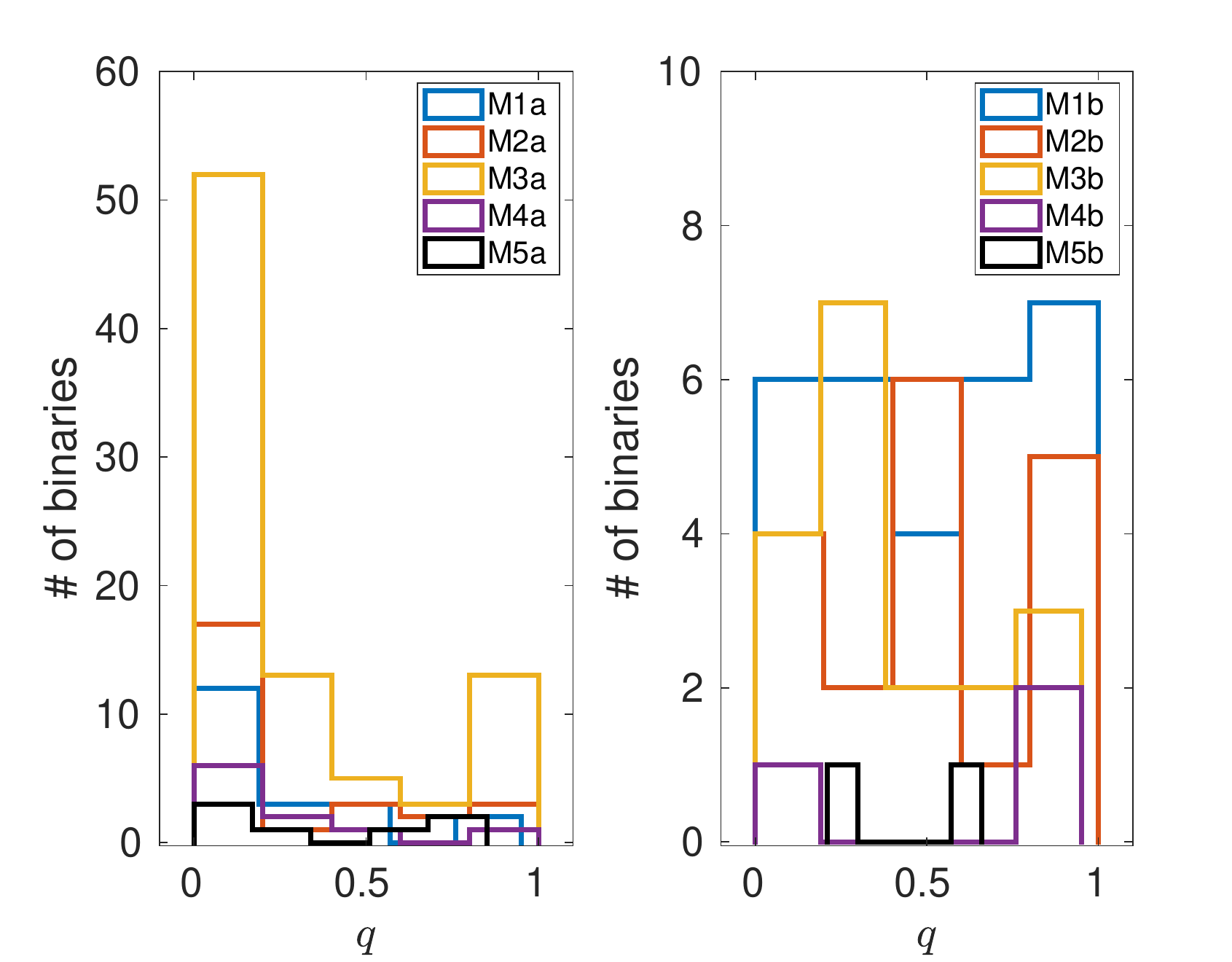}
	\caption{The top, middle and bottom panels show the distributions of the semi-major axis ($a$) in au, eccentricity ($e$), and mass ratio ($q$), respectively at the end of the simulations when the SFE reaches $\xi$ = 2 per cent. The left and right columns in each panel indicate models M1a$-$M5a and M1b$-$M5b, respectively. }
	\label{fig:figur6}
\end{figure}

The protobinary systems in our models M1a$-$M5a and M1b$-$M5b cover the full range of eccentricities $e$ = 0 $-$ 1. The initial turbulence as well as the nature of the cooling mechanism seem to have no effect on the eccentricities of the binary systems formed in our simulations. Nonetheless, the peak of the distribution of $e$ provides hints that primordial gas environments, in general, favor the formation of more eccentric binaries. This is consistent, at least qualitatively, with previous findings as \citet{kowalska2012gravitational} have reported that a large fraction of Pop. III binaries will be eccentric. However, for a large sample of 462696 of compact binaries, \citet{kowalska2012gravitational}  have shown using a binary population synthesis code that over 60 per cent of all binaries have eccentricities greater than 0.1. In this work, we have a relatively small number of binary systems i.e. 56 and 39 in model sets M1a$-$M5a and M1b$-$M5b, respectively. In the first set of models, 91 per cent of the binary systems have eccentricities above 0.1, and in the second set of models 100 per cent of the binary systems show eccentricities greater than 0.1.  \citet{belczynski2017likelihood} have combined the initial evolutionary conditions specific to Pop. III stars, as derived from N-body simulations of binary formation in primordial haloes, with Pop. III stars binary evolution models. They also determine the eccentricity range to be 0.1 $-$ 1.0 and 0.04 $-$0.99. These ranges of the eccentricities are consistent with our reported Pop. III protobinary systems. A similar range of eccentricities of 0.1 $-$ 1.0 has also been found by \citet{liu2021dynamical} using N-body simulations to study the dynamical evolution of Pop. III stellar systems and the resulting binary statistics. The MMPBs in our simulations cover a range of eccentricities 0.16 $\leq$~$e$~$\leq$ 0.88 (see Table 4). We consider the MMPBs with $e$ $\leq$ 0.7 as relatively stable systems against any orbital decay. These eccentric Pop. III protobinary systems are of great significance to understand gravitational wave spectra from compact object binaries \citep{peters1963gravitational, peters1964gravitational, damour2004phasing, brown2010effect}.  

The mass ratio $q$ associated with the Pop. III protobinaries in our models covers the full range of $q$ = 0 $-$ 1. This indicates that the formation of binaries with extreme mass ratios in the primordial universe is quite plausible, which should be detectable in the future \citep{brzozowski2022lisa}. The wide distribution of $q$ obtained in our simulations is consistent with the recent study of binary statistics of Pop. III stars performed by \citet{liu2021dynamical}. Neither the turbulence in the gas nor the  cooling mechanism plays any role in defining the distribution of $q$ in our models. The MMPBs in our models M1a $-$ M5a and M1b $-$ M5b are generally formed as unequal mass binary systems and the range covers 0.18 $\leq$~$q$~$\leq$ 0.91 (see Table 4). We note that the eccentricities and mass ratios reported in this work are subject to changes due to N-body dynamics since the simulations we performed do not predict the final distribution of the binaries. Moreover, the small mass ratios for most binaries, especially in the model set M1a$-$M5a, might be an artefact of a protostar just being born around another massive star at SFE = 2 per cent in our simulations.  
\citet{stacy2010first} have discussed a possible direct primordial formation pathway of carbon-enhanced metal-poor stars (CEMP), provided the binary components do not grow too massive. The idea is primarily based on the work by \citet{suda2004he} who suggested that if one of the components of a Pop. III binary system was an intermediate-mass star with 1 M$_{\odot}$ $\leq$ $M_{\ast}$ $\leq$ 8 M$_{\odot}$, then  the other component might undergo carbon enhancement via binary mass transfer. With the exception of the MMPB in our model M2b, we generally found no such unequal mass binaries in which a component exists in the mass range where it can become a CEMP star.

\subsection{Caveats}
We have performed hydrodynamical simulations, which
lack radiative transfer. The formation of Pop. III stars primarily depends on H$_{2}$ cooling. Its dissociation via radiation from even distant sources can suppress the star-formation process \Citep{dekel1987physical, machacek2001simulations, johnson2007local, wise2007suppression, o2008population}. \citet{stacy2007impact} have shown that a possible cosmic-ray (CR) background generated during the
first supernova explosions can affect the period of massive Pop. III star formation. We are not in a position to address the possibility of feedback from the stars which may reduce or even halt the accretion from the metal and dust-free envelope  \Citep{omukai2001formation, omukai2002upper, omukai2003formation, hosokawa2011protostellar, hosokawa2016formation, jaura2022trapping}. It is therefore of important to include such effects in future studies to further quantify their effect on binary systems. Howevere, ionisation feedback may become significant only in more slowly fragmenting cases of the collapsing primordial gas clouds, which form fewer fragments and where accretion luminosity becomes effective \citep{smith2010effects}.

Our simulations also do not include the effect of the magnetic field. While initially in a primordial gas environment, it is expected that initially, magnetic field strength is weak \citep{ando2010generation, naoz2013generation}, the small-scale dynamo may amplify the magnetic field very significantly \citep{schleicher2010small, sur2010generation, schober2012small, latif2014magnetic}. Once the accretion disc starts to develop around protostars, magnetic field lines can channel the stream of material and may drive jets and outflows from the protostars
\citep{machida2013formation, latif2016magnetic}. \citet{silk2006first} discussed the magneto-rotational instability (MRI) that can operate in the first protostellar systems. They further suggested that such effects allow primordial star formation to occur at essentially any metallicity by regulating angular momentum transfer, fragmentation, accretion and feedback in a similar manner as occurs in present-day molecular clouds. More recently, \citet{sharda2021magnetic} have performed magnetohydrodynamics simulations and found that in well-resolved simulations of the Jeans scale during the collapse, even initially weak magnetic fields  can grow exponentially via the small-scale turbulent dynamo and the large-scale mean-field dynamo to become dynamically important.

The primordial clouds we simulate are idealised and do not begin from cosmological initial conditions. It is also worth noting that deriving the final fate of the binaries requires further N-body simulations as performed by \citet{liu2021dynamical}. 

\section{Conclusions and outlook}
We present a comparison between the characteristics of collapsing primordial gas clouds and the properties of the resulting Pop. III protobinary systems that form in models considering only H$_{2}$ line cooling as well as models including H$_{2}$ line cooling followed by CIE cooling. We compare the simulations when the SFE reaches $\xi$ = 2 per cent to ensure a comparable evolutionary stage. We also explore the dependence on the turbulent properties of the gas in both types of simulations and its impact on the resulting gas morphology, the nature of Pop. III fragmentation and the orbital properties of the protobinary systems. 

Our findings suggest that turbulence in the primordial gas plays a relevant role in forming structures via networks of shocks that develop inside the collapsing gas cloud and act as birthplaces of Pop. III protostars \citep{padoan2007two}. The models which include CIE cooling show a decreasing trend concerning the number of fragments with increasing turbulent Mach numbers. The final number of protostars $N_{\rm proto}$, which is primarily dependent on protostellar merger events during the dynamical evolution of the system, nonetheless appears as rather insensitive to the initial level of subsonic turbulence as well as the specific cooling mechanism. 

For the global disc morphology at $\xi$ = 2 per cent, we find that  an increase in turbulence affects the global disc structure and causes the disc to appear misaligned with the rotational axis of the natal gas cloud, regardless of the type of cooling  we follow in our models. H$_{2}$ line cooling acting as the only gas cooling mechanism results in a more extended disc structure. We suspect that the more compact structure of the global disc in models M1b$-$M5b is related to the higher gas densities $\geq10^{14}$~cm$^{-3}$ up to which the gas cloud is allowed to collapse for the gas to undergo the formation of the super-molecules which subsequently lead to the CIE cooling. The orientation of the Pop. III protostar cluster seems to follow the orientation of the gaseous disc (even when the disc appears misaligned with respect to the rotational axis of the cloud) as the initial turbulence varies in the range of $\mathcal{M}$ = 0.1 $-$ 0.4. However, in the mildly and also the transonic turbulent models in simulations M4a, M5a, M4b, M5b, the initially misaligned disc structure is disrupted, and no final disc structure survives. The distribution of the protostars then appears to be more scattered across the collapsed part of the primordial gas cloud. A similar turbulence-dependent behavior at the scale of the circumstellar disc has been reported previously but for simulations which were performed with magnetic fields \citep{lewis2018shaken}. It will be interesting to perform MHD simulations for the primordial gas models with various levels of subsonic turbulence to explore the behaviour of the global disc structure during collapse.   

The mass accretion activity of the MMPBs shows a dependence on the cooling mechanisms that plays a role during the collapse of the gas. The MMPBs that evolve under CIE cooling exhibit more than an order of magnitude higher peak accretion rates compared to MMPBs that evolve in models with only H$_{2}$ line cooling. The mean accretion rate $\dot M_{\rm mean}$ remains of the order of $10^{-1}$ $-$ $10^{-2}$ M$_{\odot}$ yr$^{-1}$ for the most massive binary components formed in our simulations.

Cooling via the super-molecules provides a channel for the formation of more compact Pop. III binary systems. However, other orbital parameters of the first binary stars such as eccentricity $e$ and mass ratio $q$ cover the full possible range from 0 $-$ 1 and are independent of both the initial turbulence of the  gas as well as the cooling mechanism in the collapsing gas cloud. Moreover, eccentric binaries are found exhibiting rich temporal accretion signals, showing distinct waveforms, pulse structures, and duty cycles in different eccentricity regimes \citep{zrake2021equilibrium}. We believe that these eccentricity-related characteristics can be explored for the majority of eccentric Pop. III binary systems we report in this work.   

\section*{Acknowledgements}

This research was partially supported by the supercomputing infrastructure of the NLHPC (ECM$-$02). The authors acknowledge the Kultrun Astronomy Hybrid Cluster (projects ANID Programa de Astronomia Fondo Quimal QUIMAL 170001, ANID PIA ACT172033, and Fondecyt Iniciacion 11170268) for providing HPC resources that have contributed to the research results reported in this paper. Also, the Geryon cluster at the Centro de Astro-Ingenieria UC was partially used for the calculations performed in this paper. BASAL CATA PFB-06, the Anillo ACT-86, FONDEQUIP AIC-57, and QUIMAL 130008 provided funding for several improvements to the Geryon cluster. 

RR remains thankful for funding through Agencia Nacional de Investigaci\'on y Desarrollo (ANID) (project code SA77210037).

DRGS gratefully acknowledges support by the ANID BASAL projects ACE210002 and FB210003, as well as via the Millenium Nucleus NCN19-058 (TITANs). DRGS thanks for funding via Fondecyt Regular (project code 1201280).

SV wishes to thank Prof. Dr. R. Keppens and Prof.  Dr. S. Poedts for providing access to the KUL supercomputing cluster Thinking while developing and testing the code that was used in this work. He also gratefully acknowledges the support of the KUL HPC team and the NLHPC (ECM$-$02). 

RSK acknowledges financial support from the German Research Foundation (DFG) via the collaborative research centre (SFB 881, Project-ID 138713538) ``The Milky Way System'' (subprojects B1, B2, and B8), from the Heidelberg cluster of excellence EXC 2181 (Project-ID 390900948) ``STRUCTURES: A unifying approach to emergent phenomena in the physical world, mathematics, and complex data'' funded by the German Excellence Strategy, and from the European Research Council in the ERC Synergy Grant ``ECOGAL: Understanding our Galactic ecosystem -- From the disc of the Milky Way to the formation sites of stars and planets'' (grant 855130).



\section*{Data Availability}
The data underlying this article will be shared on reasonable request to the corresponding author.



\bibliographystyle{mnras}
\bibliography{example} 








\bsp	
\label{lastpage}
\end{document}